\documentstyle[seceq,mbf,epsf,wrapfig]{ptptex}
\def \beq{\begin{equation}}
\def \eeq{\end{equation}}
\def \ben{\begin{eqnarray}}
\def \een{\end{eqnarray}}
\def \bea{\begin{array}}
\def \eea{\end{array}}
\def \bem{\begin{displaymath}}  %equation without number
\def \eem{\end{displaymath}}

\def \frac#1#2{ { #1 \over #2} }

\def \ket#1{| #1 \rangle}
\def \braketa#1{\langle #1 \rangle}
\def \braketb#1#2{\langle #1 | #2 \rangle}
\def \braketc#1#2#3{\langle #1 | #2 | #3 \rangle}
\def \gtsim{\mathrel{\hbox{\raise0.2ex
     \hbox{$>$}\kern-0.75em\raise-0.9ex\hbox{$\sim$}}}}
\def \ltsim{\mathrel{\hbox{\raise0.2ex
     \hbox{$<$}\kern-0.75em\raise-0.9ex\hbox{$\sim$}}}}

\def \cJ{{ \cal J}}
\def \cH{{ \cal H}}
\def \cQ{{ \cal Q}}
\def \cB{{ \cal B}}
\def \cF{{ \cal F}}
\def \cG{{ \cal G}}

\def \cU{{ \cal U}}
\def \data{{ \rm data}}
\def \Nils{{ \rm Nils}}
\def \self{{ \rm self}}
\def \rexp{{ \rm exp}}
\def \rcal{{ \rm cal}}
\def \rms{{ \rm rms}}
\def \osc{{ \rm osc}}
\def \rot{{ \rm rot}}
\def \RPA{{ \rm RPA}}

\def \intr{{ \rm intr}}
\def \evod{{ {\rm e}\hbox{\raise-0.15ex\hbox{\rm -}}{\rm o} }}
\def \oneqp{{ 1\hbox{\raise-0.15ex\hbox{\rm -}}{\rm q.p.} }}
\def \nmax{{ n_{\rm max}}}
\def \colon{{ \mbox{:} }}
\def \maru#1{ {\mathop{#1}^{\circ}} }
\def \T{{ \rm T}}
\notypesetlogo  %comment in if to eliminate PTPTeX logo
\title{ 
Diabatic Mean-Field Description of Rotational Bands\\
in Terms of the Selfconsistent Collective Coordinate Method
\thanks{
To appear in Progress of Theoretical Physics Supplement
{\it ``Selected Topics in Boson Mapping and Time-Dependent
Hartree-Fock Methods.''}}
}
\author{
      Yoshifumi R. {\sc Shimizu} and
      Kenichi {\sc Matsuyanagi}$^{*}$
}

\inst{
 Department of Physics, Kyushu University, Fukuoka 812-8581 \\
 $^{*}$Department of Physics, Graduate School of Science,\\ 
       Kyoto University, Kyoto 606-8502
 } 

\abst{

  Diabatic description of rotational bands provides a clear-cut
picture for understanding the back-bending phenomena,
where the internal structure of the yrast band changes dramatically
as a function of angular momentum.
A microscopic framework to obtain the diabatic bands
within the mean-field approximation is presented by making use of
the selfconsistent collective coordinate method.
Applying the framework, both the ground state
rotational bands and the Stockholm bands are studied systematically
for the rare-earth deformed nuclei.
An overall agreement has been achieved
between the calculated and observed rotational spectra.
It is also shown that the inclusion of
the double-stretched quadrupole-pairing interaction
is crucial to obtain an overall agreement 
for the even-odd mass differences and
the rotational spectra simultaneously.
}

\begin{document}

\maketitle

% ***********************************
% * SECTION 1
% ***********************************
\section{ Introduction }

  Back-bending of the yrast rotational bands is one of
the most striking phenomena in the spectroscopic studies of
rapidly rotating nuclei.\cite{Hama85,Szy83} \ 
The first back-bending, which has been observed systematically
in the rotational bands of the rare-earth nuclei,
has been understood as a band-crossing
between the ground state rotational band ($g$-band)
and the lowest two quasineutron excited band ($s$-band).
A simple approach to describe the band-crossing
is the cranked mean-field approximation, where the concept of
independent particle motion in the rotating frame is fully employed.
As long as the conventional (adiabatic) cranking model is used,
however, the two bands mix at the same rotational frequency and, 
in the crossing region, loose their identities 
as individual rotational bands.
It should be noted that the difficulty lies in the fact that
the angular momenta of two bands are considerably different
in the vicinity of the crossing frequency where the mixing takes
place, especially in the case of sharp back-bendings, 
and such a mixing is largely unphysical.\cite{Hama76,MG78,BF79a}

  A key to solve this problem is to construct {\it diabatic rotational bands},
where the internal structure of the band does not 
change abruptly.\cite{Fra82,SM83,Ben89,MDV97} \ 
Once reliable diabatic bands are obtained it is rather straightforward
to mix them if the number of independent bands are few as in the case
of the first back-bending.  
Note, however, that it is highly non-trivial how to construct
reliable diabatic bands in the mean-field approximation,
because it is based on the variational principle and the mixing
at the same rotational frequency is inevitable for states with
the same quantum numbers (in the intrinsic frame).
On the other hand, in the mean-field approximation, 
the effects of rotational motion on the internal structure of the
$g$-band can be nicely
taken into account as selfconsistent changes of the deformation
and the pairing gap parameters.
Furthermore, rotation alignment effects of the quasiparticle angular
momenta are described in a simple and clear way. 
Therefore, it is desired to develop a method to describe
the rotational band diabatically within the mean-field approximation.

  In this paper, we present a powerful method to obtain
reliable diabatic rotational bands by making use of
the selfconsistent collective coordinate (SCC) method.\cite{MMSK80} \ 
The method is applied to the $g$- and $s$-bands and the results
for nuclei in the rare-earth region are compared
systematically with experimental data.
In order to reproduce the rotational spectra,
the choice of residual interaction is essential.
We use the pairing-plus-quadrupole force type interaction.\cite{BS69} \ 
However, it has been well-known that the moment of inertia 
is generally underestimated by about 20$-$30\%
if only the monopole-pairing interaction is included.\cite{NP61} \ 
Therefore, we exploit the monopole and quadrupole type interaction
in the pairing channel, and investigate the best form of
the quadrupole-pairing part.  This is done in \S2.
After fixing the suitable residual interaction, we present in \S3
a formulation to describe the diabtic rotational bands and
results of its application to nuclei in the rare-earth region.
In practical applications it often happens that a complete set
of the diabatic quasiparticle basis is necessary;
for example, in order to go beyond the mean-field approximation.
For this purpose, 
we present in \S4 a practical method to construct
the diabatic quasiparticle basis 
satisfying the orthonormality condition.
Concluding remarks are given in \S5.

% ***********************************
% * SECTION 2
% ***********************************
\section{ Quadrupole-pairing interaction suitable for deformed nuclei }
\label{sec:QPairInt}

   In this section we try to fix the form of residual interactions,
which is suitable to describe the properties of deformed rotating nuclei.
It might be desirable to use effective interactions 
like Skyrme-type interactions,\cite{VB72} \ 
but that is out of scope of the present investigation.
We assume the separable-type schematic interactions instead, and try to
fix their forms and strengths by a global fit of the basic properties;
the even-odd mass difference and the moment of inertia.

\subsection{ Residual interactions }
\label{sec:ResInt}

  The residual interaction we use in the present work
is of the following form:
\beq
  V = - G_0 P_{00}^\dagger P_{00} - G_2 \sum_K P_{2K}^\dagger P_{2K}
       -\frac{1}{2}\, \sum_K \kappa_{2K} Q_{2K}^\dagger Q_{2K},
\label{eq:resint}
\eeq
where the first and the second terms are the monopole- and
quadrupole-pairing interactions,
while the third term is the quadrupole particle-hole type interaction.
The pairing interactions are set up for neutrons and protons separately
(the $T=1$ and $T_z=\pm 1$ pairing) as usual, although it is not
stated explicitly, and only the isoscalar part is considered
for the quadrupole interaction.
The quadrupole-pairing interaction is included for purpose of
better description of moment of inertia: It has been known for
many years\cite{NP61} that the cranking moments of inertia
evaluated taking into account
only the monopole-pairing interaction underestimate the experimental
ones systematically in the rare-earth region,
as long as the monopole-pairing strength is fixed to
reproduces the even-odd mass differences.
It should be mentioned that the treatment of residual interactions
in the pairing and the particle-hole channels are different:
In the pairing channel the mean-field (pairing gap) is determined
by the interaction selfconsistently, while that in the particle-hole
channel (spatial deformation) is obtained by the Nilsson-Strutinsky
method\cite{Brack72,RNS78} and the interaction in this channel only describes
the dynamical effects, i.e. the fluctuations around the equilibrium mean-field.

   The basic quantity for deformed nuclei is
the equilibrium deformation.  For the present investigation,
where the properties of deformed rotational nuclei are systematically studied,
the Nilsson-Strutinsky method is most suitable to determine
the equilibrium deformations, because there is no adjustable parameters.
As emphasized by Kishimoto and Sakamoto,\cite{SK89} \ 
the particle-hole type quadrupole interaction for deformed nuclei
should be of the double-stretched form:\cite{SK89,Kis75,Mar84}
\beq
    Q_{2K} = \sum_{ij}q_{2K}(ij)\, c_i^\dagger c_j, \quad
      q_{2K}(ij) = \braketc{i}{( r^2 Y_{2K} )''}{j},
\label{eq:dblQ}
\eeq
where $c_i^\dagger$ is the nucleon creation operator in the
Nilsson state $\ket{i}$.
$(O)''$ means that the Cartesian coordinate in the operator $O$ 
should be replaced such as
$x_k \rightarrow x''_k \equiv (\omega_k/\omega_0)x_k$ ($k=x,y,z$),
where $\omega_x$, $\omega_y$ and $\omega_z$ are
frequencies of the anisotropic oscillator potential
and related to the deformation parameter
$(\epsilon_2,\gamma)$;\cite{RNS78,NR95} \ 
here $\hbar\omega_0 \equiv \hbar(\omega_x\omega_y\omega_z)^{1/3}
=41.0/A^{1/3}$ MeV ($A$ is the mass number).
Then the selfconsistent condition gives, at the equilibrium shape,
a vanishing mean value
for the double-stretched quadrupole operator, 
$\braketa{ Q_{2K} } = 0$, and thus the meaning
of residual interaction is apparent for the double-stretched interaction.
Moreover, the force strengths are determined at the same time
to be the so-called selfconsistent value,
\beq
  \kappa_{2K}=\kappa_2^\self
  =\frac{4\pi}{3} \frac{\hbar \omega_0}{AR^2_0 b^2_0},
    \quad {\rm with} \quad
  b^2_0= \frac{\hbar}{M\omega_0},\;\;
  R_0=1.2 A^{1/3}\;{\rm fm},
\label{eq:kQself}
\eeq
by which the $\beta$- and $\gamma$-vibrational excitations are
correctly described.
Strictly speaking, the vanishing mean value of
$Q_{2K}$ holds only for the harmonic oscillator model.
It is, however, easily confirmed that the mean value vanishes
in a good approximation in the case of Nilsson potential.
In fact the calculated ratio of mean values of
the double-stretched and non-stretched quadrupole operator is
typically within few percent,
if the deformation parameter determined by the Strutinsky procedure is used.

  Pairing correlations are important
for the nuclear structure problem as well.  The operators entering
in the pairing type residual interactions are of the form
\beq
  P_{00}^\dagger = \sum_{i>0} c_i^\dagger c_{\tilde i}^\dagger, \quad
  P_{2K}^\dagger = \sum_{ij>0} p_{2K}(ij)\, c_i^\dagger c_{\tilde j}^\dagger,
\label{eq:PairOP}
\eeq
where ${\tilde j}$ denotes the time-reversal conjugate
of the Nilsson state $j$.
In contrast to the residual interactions in the particle-hole channel,
there is no such selfconsitency condition known in the pairing channel.
Therefore, we use the Hartree-Bogoliubov (HB)
procedure (exchange terms are neglected) for the pairing interactions,
only the monopole part of which leads to the ordinary BCS treatment.
Note that the generalized Bogoliubov
transformation is necessary in order to treat 
the quadrupole-pairing interaction,
since the pairing potential becomes state-dependent
and contains non-diagonal elements:
\beq
    \Delta_{ij} = \Delta_{00}\,\delta_{ij}
      + \sum_K \Delta_{2K}\, p_{2K}(ij),
\label{eq:QpairPot}
\eeq
where
$\Delta_{00} = G_0 \,\braketa{P_{00}}$ and
$\Delta_{2K} = G_2 \,\braketa{P_{2K}}$, the expectation values being
taken with respect to the resultant HB state.

   For the application of these residual interaction we are mainly
concerned with deformed nuclei in the rare-earth region,
where the neutron and proton numbers are considerably different.
In such a case, the ``iso-stretching'' of
multipole operators,\cite{BK68,Sak93} \ 
$Q_\tau \rightarrow (2N_\tau/A)^{2/3}\,Q_\tau$ for $\tau=\nu,\pi$
($N_\tau$ denotes the neutron or proton number and $A$ the mass number),
is necessary in accordance with the difference
of the oscillator frequencies, $\omega_0^\tau = (2N_\tau/A)^{1/3} \omega_0$,
or of the oscillator length, $(b_0^\tau)^2 = (2N_\tau/A)^{-1/3} b^2_0$.
We employ this modification for
the quadrupole interaction in the particle-hole channel.

\subsection{ Treatment of pairing interactions }
\label{sec:QPair}

   As for the quadrupole-pairing part, there are at least three variants
that have been used
in the literature.\cite{Hama74,Dieb84,Garr82,SK90,KSKK96,SW94} \ 
Namely, they are non-stretched, single-stretched and double-stretched
quadrupole-pairing interactions, where the pairing form factor
in the operator in Eq.~(\ref{eq:PairOP}) is defined as,
\beq
  p_{2K}(ij) = \braketc{i}{r^2Y_{2K}}{j}, \,\,\,
  \braketc{i}{(r^2Y_{2K})'}{j}, \,\,\, \braketc{i}{(r^2Y_{2K})''}{j},
\label{eq:PQform}
\eeq
respectively.  The single-stretching of operators is analogously
performed by the replacement,
$x_k \rightarrow x'_k \equiv \sqrt{\omega_k/\omega_0}~x_k$ ($k=x,y,z$).
Note that there are matrix elements between the Nilsson states 
with $\Delta N_\osc=\pm 2$ in Eq.~(\ref{eq:PQform}).
We have neglected them in the generalized Bogoliubov transformation
in accordance with the treatment of the Nilsson potential, which is
arranged to have vanishing matrix elements
of $\Delta N_\osc=\pm 2$.\footnote{
 The hexadecapole deformation leads extra $\Delta N_\osc=\pm 2,\, \pm 4$
 coupling terms, but they are neglected in our calculations.
}

   Consistently to the Nilsson-Strutinsky method, we use
the smoothed pairing gap method\cite{Brack72} in which the monopole-pairing
force strength is determined for a given set of
single-particle energies by
\beq
    \frac{2}{G_0}= \frac{1}{2}\,{\widetilde g}_{\rm F}
      \log{\Bigl(\Lambda/{\widetilde \Delta}+
      \sqrt{ (\Lambda/{\widetilde \Delta})^2 +1 }\,\Bigr)},
\label{eq:smG0}
\eeq
where ${\widetilde g}_{\rm F}$ is the Strutinsky smoothed single-particle
level density at the Fermi surface, $\Lambda$ is the cutoff energy
of pairing model space, for which we use $\Lambda=1.2\hbar\omega_0$,
and ${\widetilde \Delta}$ is the smoothed pairing gap.
We introduce a parameter $d$ (MeV) to control the strength of 
the monopole-pairing force by
\beq
    {\widetilde \Delta} = \frac{d}{\sqrt{A}},
\label{eq:smDelta}
\eeq
through Eq.~(\ref{eq:smG0}), where the same smoothed pairing gap is used
for both neutrons and protons, for simplicity.
As for the quadrupole-pairing force strength, we take the following form,
\beq
   G_2 = G_0 \,\frac{g_2}{R^4_0}, \quad
   {\rm with}\quad R_0=1.2A^{1/3}\;{\rm fm}.
\label{eq:G2byG0}
\eeq
Thus, we have two parameters $d$ (in MeV) and $g_2$ for the residual
interactions in the pairing channel.

    It is worthwhile mentioning that Eq.~(\ref{eq:smG0}) gives the form,
\beq
  G^\tau_0 \approx \frac{c}{A} \Bigl(\frac{2N_\tau}{A}\Bigr)^{-1/3},
   \quad {\rm with}\quad c=\frac{41.0}{3^{2/3}2^{4/3}}
      \log{\bigl(2\Lambda/{\widetilde \Delta}\bigr)},
\label{eq:G0HO}
\eeq
for the semiclassical treatment of the isotropic harmonic
oscillator model,\cite{BMtextI,BMtextII} \ 
where ${\widetilde g}_{\rm F}^\tau \approx (3N)^{2/3}/(\hbar \omega_0^\tau)$,
and it is a good approximation for the Nilsson potential.
The quantity $\log{\bigl(2\Lambda/{\widetilde \Delta}\bigr)}$
depends very slowly on the mass number and can be replaced by
a representative value for a restricted region of mass table.
Taking $A=170$, one obtains $c \approx 23$, which gives
the monopole-pairing force strength often used
for nuclei in the rare-earth region.

\subsection{ Determination of parameters $d$ and $g_2$ }
\label{sec:detQPair}

    Now let us determine the form of the quadrupole-pairing interaction.
Namely, we would like to answer the question of which form factor
in Eq.~(\ref{eq:PQform}) is best,
and of what are the values of the parameters, $d$ and $g_2$, introduced
in the previous subsection.  For this purpose, we adopt the following
criteria; the moments of inertia $\cJ_0$ of the Harris formula\cite{Har65}
and the even-odd mass differences (the third order formula\cite{BMtextI})
for even-even nuclei should be simultaneously reproduced
as good as possible.
Since the neutron contribution
is more important for the moment of inertia, we have
used the even-odd mass difference for neutrons, $E_\nu^{(\evod)}$.
Then it turns out that the proton even-odd mass difference is also
reasonably well reproduced as long as the same smoothed pairing gap
is used for neutrons and protons.
Thus, the two parameters $d$ (MeV) and $g_2$ are searched so as to
minimize the root-mean-square deviations of these quantities divided by
their average values,
\beq
   X_\rms(x) = \Bigl[ \frac{1}{N_\data} \sum_{i=1}^{N_\data}
      \bigl( x_i^{(\rexp)} - x_i^{(\rcal)} \bigr)^2 \Bigr]^{1/2}
  \Big/\, \Bigl[ \frac{1}{N_\data} \sum_{i=1}^{N_\data} x_i^{(\rexp)} \Bigr]
\label{eq:RMS}
\eeq
for $x=\cJ_0$ and $E_\nu^{(\evod)}$.
Nuclei used in the search are chosen
from even-even rare-earth nuclei in Table~\ref{tab:Nucls},
thus $N_\data=83$~(58) for $x=E_\nu^{(\evod)}$~($\cJ_0$).

%%%%%%%%%%%
%  Table  %
%%%%%%%%%%%

\begin{table}
\caption{
 Nuclei included for the search of the pairing interaction parameters,
 $d$ and $g_2$.
  }
\label{tab:Nucls}
\begin{center}
\begin{tabular}{lcccccc}
\hline\hline
 & $_{64}$Gd & $_{66}$Dy & $_{68}$Er & $_{70}$Yb & $_{72}$Hf & $_{74}$W \\
\hline
$N$ for $E_\nu^{(\evod)}$ &
 76$-$100 & 78$-$102 & 80$-$104 & 80$-$108 & 84$-$110 & 86$-$114 \\
$N$ for $\cJ_0$ &
 86$-$96 & 86$-$100 & 86$-$102 & 86$-$108 & 90$-$110 & 92$-$114 \\
\hline
\end{tabular}
\end{center}
\end{table}

% **********
% * Figure
% **********
\begin{figure}
\epsfysize=8.0cm
\centerline{\epsfbox{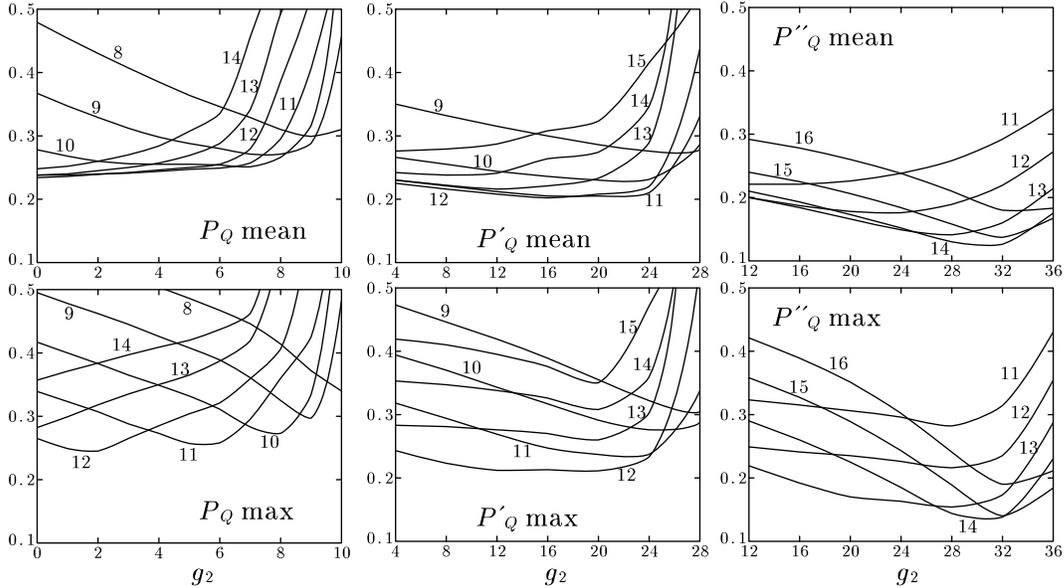}}
\caption{
 Root-mean-square deviations of neutron even-odd mass differences
 and moments of inertia, calculated by using the non-stretched (left),
 single-stretched (middle), and double-stretched (right) residual
 quadrupole-pairing interactions.  The upper panels shows the results
 for ${\overline X}_\rms$ and the lower panels
 for $X_\rms^{\rm M}$,
 see Eqs.~(\protect\ref{eq:meanRMS}) and~(\protect\ref{eq:maxRMS}).
 They are calculated as functions
 of the two parameters $d$ and $g_2$.  Each curve is drawn
 with a fixed value of $d$ (MeV), which is attached near the curve,
 as a function of $g_2$.
  }
\label{fig:rskG2}
\end{figure}

   Note that the neutron even-odd mass difference has been calculated
in the same way as the experimental data by taking
the third order difference of calculated binding energies
for even and odd $N$ nuclei,
where the blocking HB calculation has been done for odd-mass nuclei.
In the Nilsson-Strutinsky method the grid of deformation parameters
$-0.08 \le \epsilon_2 \le 0.40$ and
$-0.08 \le \epsilon_4 \le 0.12$ with an interval of $0.04$ are used.
The $ls$ and $ll$ parameters of the Nilsson potential are taken
from Ref.~\citen{BR85}.
We have assumed the axial symmetry in the calculation of
this subsection since only the ground state properties are examined.
Experimental binding energies are taken from
the 1993 Atomic Mass Evaluation.\cite{AW93} \ 
As for the experimental moment of inertia, the Harris parameters,
$\cJ_0$ and $\cJ_1$, are calculated from the observed excitation energies
of the $2^+$ and $4^+$ states belonging to the ground state band,
experimental data being taken from Ref.~\citen{Sakai84} and
the ENSDF database.\cite{ENSDF} \ 
If the value of $\cJ_0$ calculated in this way becomes negative
or $\cJ_1$ greater than 1000 $\hbar^4$/MeV$^3$
(this happens for near spherical nuclei), then
$\cJ_0$ is evaluated by only using the $2^+$ energy,
i.e. by $3/E_{2^+}$.
The Thouless-Valtion moment of inertia,\cite{RStext} \ 
which includes the effect of the $K=1$ component of the
residual quadrupole-pairing interaction,
is employed as the calculated moment of inertia.
Here, again, the matrix elements between states with $\Delta N_\osc=\pm 2$
are neglected for simplicity in the same way as in the step of 
diagonalization of the mean-field Hamiltonian.
The contributions of them are rather small for
the calculation of moment of inertia,
since the $\Delta N_\osc=\pm 2$ matrix elements of the angular momentum
operator are smaller than the $\Delta N_\osc=0$ ones
by a factor $\approx \epsilon_2$ and the energy denominators are larger.
We have checked that those effects are less than 5 \% for
the Thouless-Valatin moment of inertia in well deformed nuclei.

    In Fig.~\ref{fig:rskG2} we show 
root-mean-square deviations of the result of calculation for
neutron even-odd mass differences
and moments of inertia.  We have found that the behaviors of
these two quantities, $X_\rms(E_\nu^{(\evod)})$ and $X_\rms(\cJ_0)$,
as functions of $g_2$ with fixed $d$ are opposite, and so 
the mean value
\beq
 {\overline X}_\rms = \frac{1}{2}\,
 \bigl(X_\rms(E_\nu^{(\evod)})+X_\rms(\cJ_0) \bigr)
\label{eq:meanRMS}
\eeq
become almost constant, especially for the case of the non-stretched
quadrupole-pairing.  Therefore, we also display the results
for the maximum among the two,
\beq
  X_\rms^{\rm M} =
  {\rm max}\bigl\{ X_\rms(E_\nu^{(\evod)}),\,X_\rms(\cJ_0) \bigr\}.
\label{eq:maxRMS}
\eeq
As is clear from Fig.~\ref{fig:rskG2}, the best fit is obtained
for the double-stretched quadrupole-pairing interaction with
$d=14$ (MeV) and $g_2=30$.
It should be mentioned that the value of $g_2$ is close to the one
$g_2=28\pi/3$ in Ref.~\citen{Hama74}, where it is derived from
the multipole decomposition of the $\delta$-interaction and this argument
is equally applicable if the double-stretched coordinate is used
in the interaction.
It is interesting to notice that if the non-stretched or the
single-stretched quadrupole-pairing interaction is used, then one cannot make
either ${\overline X}_\rms$ or $X_\rms^{\rm M}$ smaller than 0.2.
${\overline X}_\rms$ in the non-stretched case
is rather flat as a function of $g_2$ and the minimum occurs at
$d=12$ (MeV) and $g_2=0$ (no quadrupole-pairing). 
$X_\rms^{\rm M}$ in the non-stretched case takes the
minimum at small quadrupole-pairing, $d=12$ (MeV) and $g_2=2$.
Both ${\overline X}_\rms$ and $X_\rms^{\rm M}$
are flat as a function of $g_2$ also in the single-stretched case,
and take the minimum at $d=12$ (MeV) and $g_2=16$.
In contrast, the double-stretched interaction gives
well developed minima for both ${\overline X}_\rms$ and $X_\rms^{\rm M}$.
These results clearly show that one has to use the double-stretched
quadrupole-pairing interaction.

% **********
% * Figure
% **********
\begin{figure}
\epsfysize=8.0cm
\centerline{\epsfbox{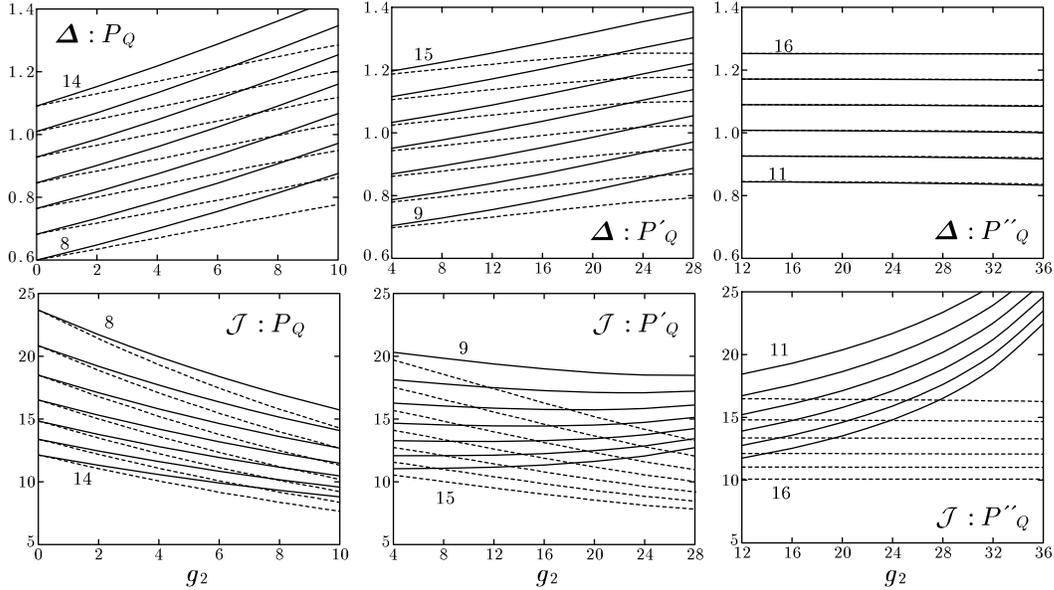}}
\caption{
 Pairing gaps (upper panels)
 and moments of inertia (lower panels), calculated
 by using the non-stretched (left), single-stretched (middle),
 and double-stretched (right) residual quadrupole-pairing interactions.
 Average pairing gaps ${\overline \Delta}$ and
 monopole-pairing gaps $\Delta_{00}$ (MeV),
 see Eq.~(\protect\ref{eq:avrGap}), are displayed by solid and dashed
 curves, respectively, in the upper panels, while
 Thouless-Valatin and Belyaev moments of inertia ($\hbar^2$/MeV)
 (c.f. Eq.~(\protect\ref{eq:THVmom}))
 are displayed as solid and dashed curves, respectively,
 in the lower panels.
 They are calculated as functions of the two parameters $d$ and $g_2$.
 Each curve is drawn with a fixed value of $d$ (MeV),
 which is attached near the curve and changed by step of 1 MeV,
 as a function of $g_2$.
 The calculation has been done for a typical deformed nucleus,
 $^{168}$Yb, with deformation
 parameters $(\epsilon_2,\,\epsilon_4)=(0.2570,\,0.0162)$.
  }
\label{fig:djG2}
\end{figure}

    One may wonder why the non- and single-stretched interactions do not
essentially improve the root-mean-square deviations.
The quadrupole-pairing interaction affects $E_\nu^{(\evod)}$ and $\cJ_0$
in two ways:
One is the static (mean-field) effect through the change of static
pairing potential~(\ref{eq:QpairPot}), and
the other is a dynamical effect (higher order than the mean-field approximation)
and typically appears as the Migdal term
in the Thouless-Valatin moment of inertia
(c.f. Eq.~(\ref{eq:THVmom}) and~(\ref{eq:THVmoma})).
The former effect can be estimated by the averaged pairing gap,
\beq
  {\overline \Delta} = \sum_i \Delta_{ii} \,/\, \sum_i 1
   = \Delta_{00} + \sum_K \Delta_{2K} \sum_i p_{2K}(ii) \,/\, \sum_i 1,
\label{eq:avrGap}
\eeq
where the summation is taken over the Nilsson basis state $i$'s
included in the pairing model space.
Stronger quadrupole-pairing interaction results in larger
${\overline \Delta}$, which leads to the increase of even-odd mass difference
on one hand and the reduction of moment of inertia on the other hand.
The Migdal term coming from the $K=1$ component of the
quadrupole-pairing interaction makes the moment of inertia larger
when the force strength is increased.
Therefore, the moment of inertia either increases
or decreases as a function of force strength,
depending on which effect is stronger.
In Fig.~\ref{fig:djG2}, we have shown the energy gap
and the moment of inertia for a typical rare-earth deformed nuclei
$^{168}$Yb as functions of the two parameters $d$ and $g_2$ in parallel
with Fig.~\ref{fig:rskG2}.
One can see that the average as well as monopole-pairing gaps
increase rapidly as functions of the quadrupole-pairing strength
if the non-stretched interaction is used.  This static effect is
so strong that the Thouless-Valatin moment of inertia decreases.
In the case of the single-stretched case, similar trend is
observed for the pairing gap, though it is not so dramatic
as in the case of non-stretched interaction.  The static effect
almost cancels out the dynamical effect and then
the Thouless-Valatin moment of inertia stays almost constant
against $g_2$ in this case.
On the other hand, if one uses the double-stretched interaction,
the pairing gap stays almost constant as a function of $g_2$.
This is because $\braketa{P_{2K}} \approx 0$ holds in a very
good approximation, which is in parallel with the fact that
the quadrupole equilibrium shape satisfies the selfconsistent
condition, $\braketa{Q_{2K}} = 0$,
for the double-stretched quadupole operator.
Thus the effect of the double-stretched quadrupole-pairing interaction
plays a similar role as the particle-hole interaction channel;
it acts as a residual interaction and does not contribute to
the static mean-field.

\subsection{ Results of calculation }
\label{sec:resCal}

    It has been found in the previous subsection that
the double-stretched form of the quadrupole-pairing interaction
with parameters $d=14$ MeV and $g_2=30$
gives best fitting for the even-odd mass differences and
the moments of inertia in the rare-earth region.
Resulting root-mean-square deviations are
$X_\rms(E_\nu^{(\evod)},\,\cJ_0)=(0.115,\,0.136)$.
If one uses $(d,g_2)=(13,28)$ or $(12,20)$, as examples,
those quantities become
$X_\rms(E_\nu^{(\evod)},\,\cJ_0)=(0.154,\,0.127)$ or $(0.235,\,0.121)$,
respectively.
Therefore, making the two quantities smaller is complementary
as discussed in \S\ref{sec:detQPair}.

% **********
% * Figure
% **********
\begin{figure}
\epsfysize=8.8cm
\centerline{\epsfbox{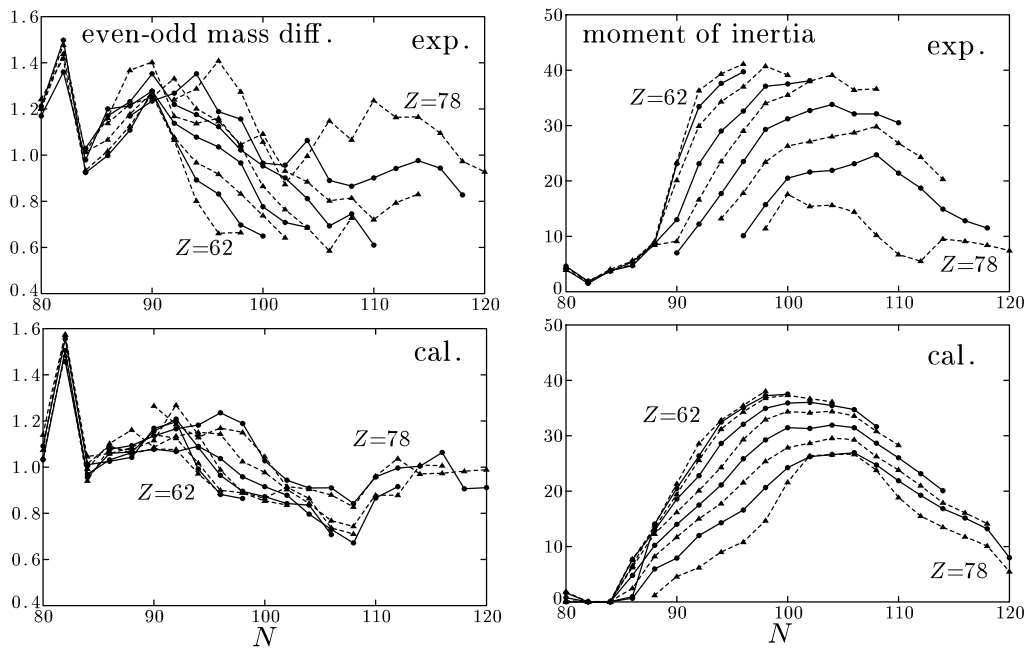}}
\caption{
 Comparison of calculated even-odd mass differences (left panels, in MeV)
 and moments of inertia (right panels, in $\hbar^2$/MeV)
 with experimental data for nuclei in the rare-earth region.
 Experimental data are displayed in the upper panels
 while the calculated ones in the lower panels.
 Isotopes with $Z=62-78$ are connected by solid ($Z=0$ mod 4)
 or dashed ($Z=2$ mod 4) curves as functions of neutron number $N$.
 The double-stretched quadrupole-pairing interaction is used
 with parameters $d=14$ MeV and $g_2=30$.
  }
\label{fig:eodmomm}
\end{figure}

% **********
% * Figure
% **********
\begin{figure}
\epsfysize=8.8cm
\centerline{\epsfbox{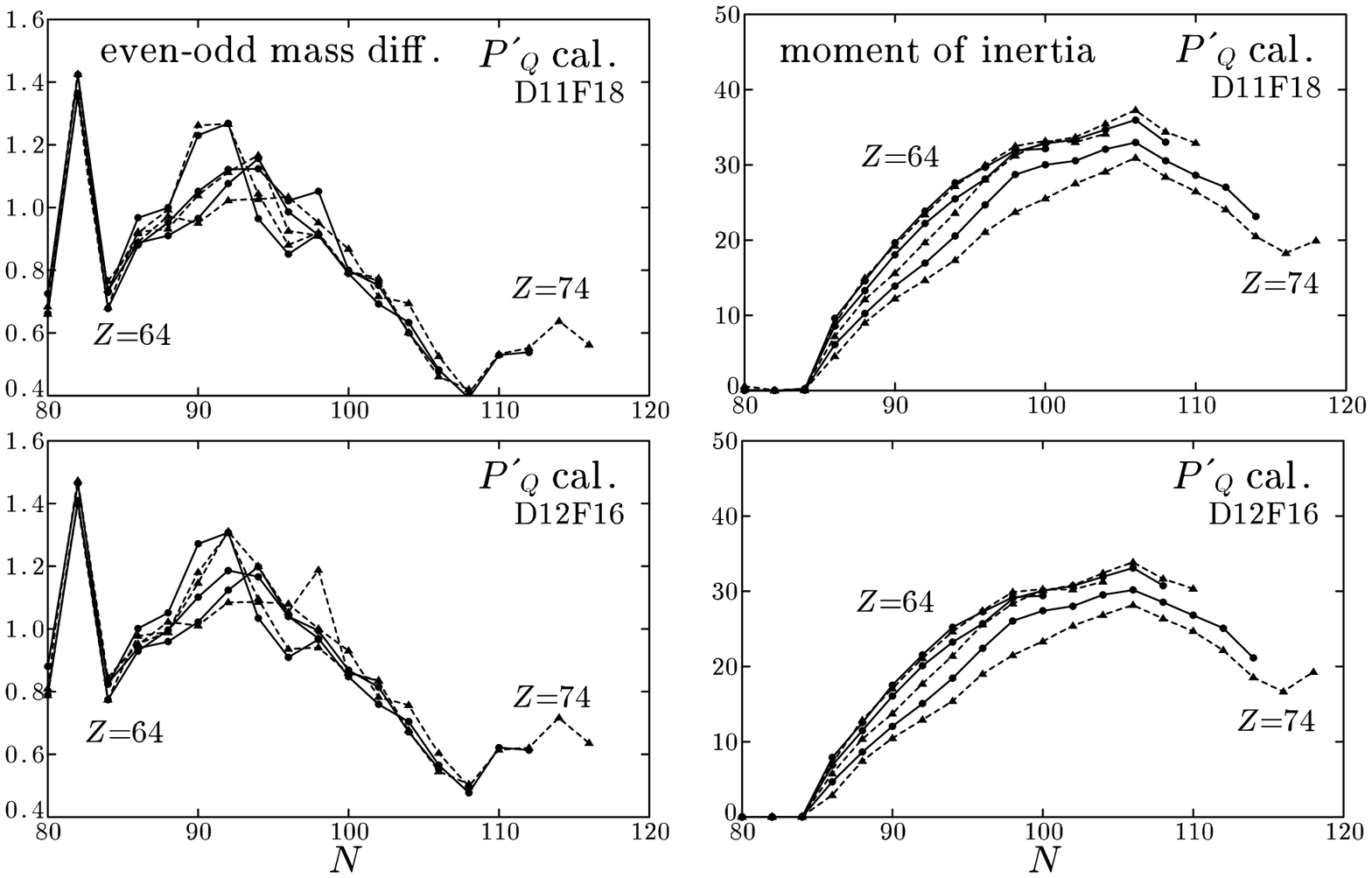}}
\epsfysize=8.8cm
\centerline{\epsfbox{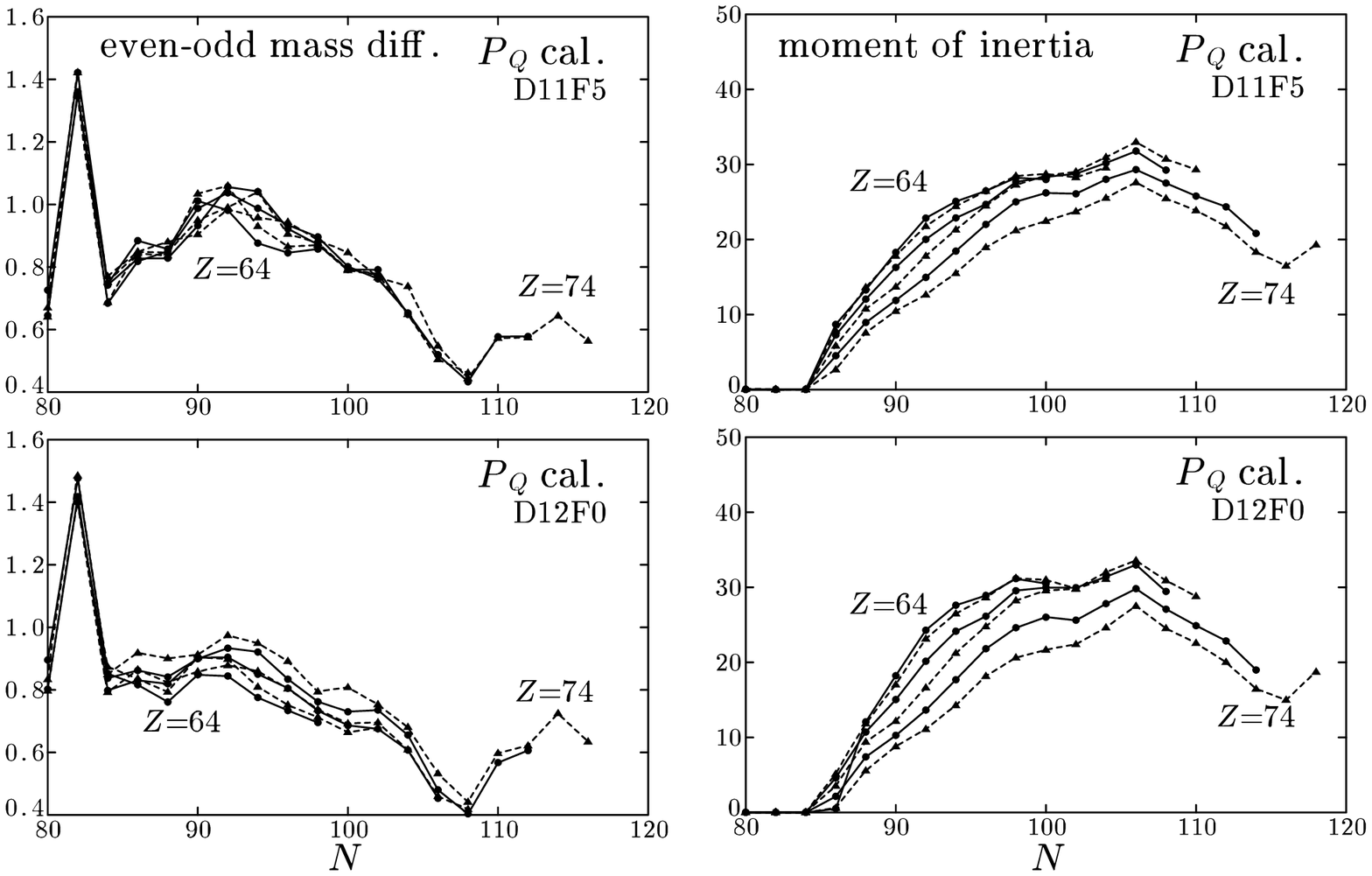}}
\caption{
 Even-odd mass differences (left panels)
 and moments of inertia (right panels) for $Z=64-74$ isotopes, calculated
 by using the single- and non-stretched quadrupole-pairing interactions.
 The panels from top to bottom show the results of the single-stretched cases
 with parameters ($d=11$ MeV, $g_2=18$) and ($d=12$ MeV, $g_2=16$), 
 and of the non-stretched cases with parameters ($d=11$ MeV, $g_2=5$)
 and ($d=12$ MeV, $g_2=0$), respectively.
  }
\label{fig:eodmom10}
\end{figure}

% **********
% * Figure
% **********
\begin{figure}
\epsfysize=8.8cm
\centerline{\epsfbox{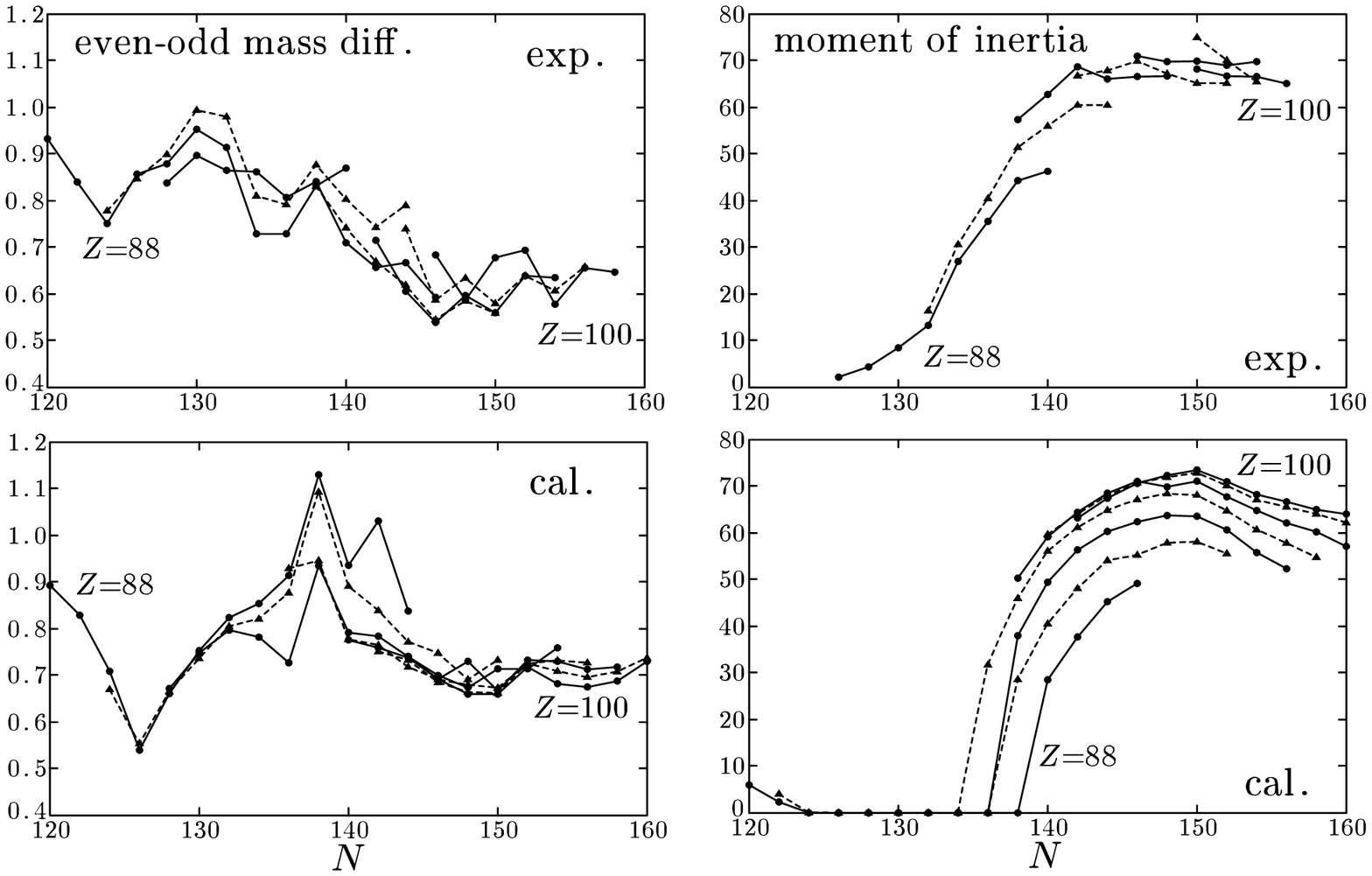}}
\caption{
 Same as Fig.~\protect\ref{fig:eodmomm}, but for nuclei in the
 actinide region.
 Isotopes with $Z=88-100$ are connected by solid ($Z=0$ mod 4)
 or dashed ($Z=2$ mod 4) curves as functions of neutron number $N$.
  }
\label{fig:eodmomh}
\end{figure}

     We compare the results of calculation with
experimental data in Fig.~\ref{fig:eodmomm}
as functions of neutron number.
In this calculation the results of Sm ($Z=62$), Os ($Z=76$) 
and Pt ($Z=78$) isotopes are also included,
which are not taken into account in the fitting procedure.
As is clear from the figure, both even-odd mass differences
and moments of inertia are not well reproduced
in heavy Os and Pt isotopes; especially
even-odd mass differences are underestimated by about 20\%,
and moments of inertia  overestimated by about up to 50\%
in Pt nuclei with $N \gtsim 100$.
In these nuclei, low-lying spectra suggest that they
are $\gamma$-unstable, and therefore correlations
in the $\gamma$ degrees of freedom are expected to play an important role.
Except for these nuclei,
the overall agreements have been achieved, particularly for
deformed nuclei with $N \approx 90-100$.
It is, however, noted that some features seen in experimental
data are not reproduced in the calculation:
(1) The maximum at $N=90$ and
the minimum at $N=106$ or $110$ in the even-odd mass difference are
shifted to $N=92$ and $N=98$, respectively.
This is because details of the neutron single-particle level spacings
in the present Nilsson potential are slightly inadequate.
(2) The proton number dependences of both the even-odd mass difference
and the moment of inertia are too weak: curves of both quantities
bunch more strongly in the calculation.  This trend is clearer
in light $Z$ nuclei, $Z \le 68$, for example, Gd or Dy;
the even-odd mass difference in these isotopes decreases more slowly
as a function of neutron number in the calculation,
which results in the slower increase of the moment of inertia.
This problem suggests that some neutron-proton correlations might be
necessary.

    For comparison sake, results obtained by using
the quadrupole-pairing interactions
of the single-stretched and the non-stretched types
are displayed in Fig.~\ref{fig:eodmom10}.
In the calculation of the single-stretched case, the values of
the two parameters, $d=11$ MeV and $g_2=18$, are employed, resulting
$X_\rms(E_\nu^{(\evod)},\,\cJ_0)=(0.240,\,0.170)$, in one case, and
the values $d=12$ MeV and $g_2=16$, resulting
$X_\rms(E_\nu^{(\evod)},\,\cJ_0)=(0.192,\,0.214)$, in another case.
Comparing with the experimental data in Fig.~\ref{fig:eodmomm},
the decrease of even-odd mass difference with neutron number is too strong,
while the increase of moment of inertia near $N \approx 90$ is too slow.
In the calculation of non-stretched case, the values of
the two parameters, $d=11$ MeV and $g_2=5$ are employed, resulting
$X_\rms(E_\nu^{(\evod)},\,\cJ_0)=(0.257,\,0.237)$, in one case, and
the values $d=12$ MeV and $g_2=0$, resulting
$X_\rms(E_\nu^{(\evod)},\,\cJ_0)=(0.265,\,0.204)$, in another case.
The average values of the even-odd mass difference are considerably
smaller and those of the moment of inertia are 20$-$30\% smaller
compared to the experimental data.  Note that the last case
($d=12$ MeV and $g_2=0$) is nothing but the calculation without
the quadrupole-pairing interaction.
The trend of weak proton number dependence does not change for all
three forms of the quadrupole-pairing interaction.

    The merit of the Nilsson-Strutinsky method is that
a global calculation is possible once the mean-field potential is given.
We have then performed the calculation for nuclei in the actinide region
with the same pairing interaction and the parameters
as in the rare-earth region, i.e.
the double-stretched quadrupole-pairing with $d=14$ MeV and $g_2=30$.
The result is shown in Fig.~\ref{fig:eodmomh}.
Nuclei in the light actinide region are spherical or weakly deformed
with possible octupole deformations.
The experimental moments of inertia suggest that
nucleus in this region begins to deform at $N \approx 134$,
and gradually increases the deformation
until a rather stable deformation is established at $N \gtsim 140$.
In the nuclei with $N=88$ and $90$, the neutron number at which
the deformation starts to grow is too large in the calculation,
and the even-odd mass differences take considerably different behaviour
from the experimental data.
This disagreement possibly suggests
the importance of octupole correlations.
Except for these difficiencies, both even-odd mass differences
and moments of inertia in heavy well-deformed nuclei
are very well reproduced in the calculation.
It should be emphasized that the parameters fixed in \S\ref{sec:detQPair}
for the rare-earth region are equally well applicable
for the actinide region.

% ***********************************
% * SECTION 3
% ***********************************
\section{ SCC method for constructing diabatic rotational bands }
\label{sec:SCCmethod}

     The SCC method\cite{MMSK80} is a theory aiming at microscopic
description of large amplitude collective motions in nuclei.
The rotational motion is one of the most typical large amplitude motions.
Therefore it is natural to apply the SCC method to the nuclear collective
rotation.  In Ref.~\citen{SM85}, this line has been put into practice
for the first time in order to obtain the diabatic rotational bands,
where the interband interaction associated with
the quasiparticle alignments is eliminated.
It has also been shown that the equation of path in the SCC method
leads to the selfconsistent cranking model
in the case of rotational motion.
Corresponding to the uniform rotation about one of the principal axes
of nuclear deformation,
the one-dimensional rotation has been considered as
in the usual cranking model.
We keep this basic feature in the present work.

     More complete formulation and its application to the ground state
rotational bands ($g$-bands) in realistic nuclei have been done
in Ref.~\citen{TMS91},
followed by further applications to the Stockholm bands
($s$-bands)\cite{Tera92}
and improved calculations with including the quadurpole-pairing
interaction.\cite{Tera94} \ 
In these works the basic equations of the SCC method have been solved
in terms of the angular momentum expansion ($I$-expansion).
Thus, the $A$ and $B$ parameters in the rotational energy expansion,
$E(I)=A\, I(I+1) + B\, [I(I+1)]^2$, have been studied in detail.
It is, however, well known that applicability of the $I$-expansion 
is limited to relatively low-spin regions.
This limitation is especially severe in the case of the $s$-bands:
One has to take the starting angular momentum $I_0$
($\approx 10\hbar$\cite{Tera92}) and the expansion in terms of
$(I-I_0)$ is not very stable.
Because of this problem comparisons with experimental data
have not been possible for the $s$-bands.\cite{Tera92} \ 
In the present study, the rotational frequency expansion is utilized
instead, according to the original work.\cite{SM85} \ 
Then the diabatic cranking model is naturally derived.
Thus, after obtaining the diabatic quasiparticle states,
we construct the $s$-band as the two quasiparticle aligned band
on the vacuum $g$-band at given rotation frequencies.
This is precisely the method of the cranked shell model,\cite{BF79} \ 
which has been established as a powerful method to understand
the high-spin rotational bands accompanying quasiparticle excitations.

    Another important difference of the present work
from Refs.~\citen{TMS91,Tera92,Tera94} is that
the expansion method based on
the normal modes of the random phase approximation (RPA)
is used for solving the basic equations in these references.  
The method is very convenient to investigate detailed contents
of the rotation-vibration couplings, e.g. how each normal mode
contributes to the rotational $A$ and/or $B$ parameters,
as has been discussed in Refs.~\citen{TMS91,Tera92}.
On the other hand, we are aiming at a systematic study
of rotational spectra of both $g$- and $s$-bands
in the rare-earth region.  Then the use of the RPA response-function 
matrix is more efficient for such a purpose,
because it is not necessary to solve the RPA equation
for all the normal modes explicitly.

    It has to be mentioned that the problem of nucleon number conservation,
i.e. the pairing rotation, can be treated similarly.\cite{Mat86} \ 
Actually, if the SCC method is applied to the spatial rotational motion,
the mean value of the nucleon number changes as 
the angular momentum or the rotational frequency increases.
A proper treatment of the pairing rotations is
required, i.e. the coupling of the spatial and pairing
rotations should be included.\cite{Mat86} \ 
However, it has been found\cite{TMS91}
that the effect of the coupling is negligibly small
for the case of the rotational motion in well deformed nuclei.
Therefore, we simply neglect the proper treatment of the nucleon number
in the following.

     Although it is not the purpose of this paper to review
applications of the SCC method to other nuclear structure phenomena,
we would here like to cite a brief review\cite{Mat92} and
some papers, in which low-frequency quadrupole
vibrations are analyzed on the basis of the SCC method:
anharmonic gamma vibrations,\cite{Mat84,MMgamma,MSMgamma} \
shape phase transitions in Sm isotopes,\cite{TYT89,YTT89,YT89} \ 
anharmonicities of the two phonon states in Ru and Se
isotopes,\cite{Aiba90} \ 
single-particle levels and configurations in the shape phase
transition regions,\cite{Yamada91} \ 
and a derivation of the Bohr-Mottelson type collective Hamiltonian and
its application to transitional Sm isotopes.\cite{Yamada93}

\subsection{ Basic formulation }
\label{sec:SCCbasic}

     The starting point of the SCC method is the following time-dependent
Hartree-Bogoliubov (TDHB) mean-field state
\beq
   \ket{\phi(\theta,I_x)}=W(\theta,I_x)\ket{\phi_0},
\label{eq:sccWFa}
\eeq
which is parametrized by the time-dependent collective 
variables $\theta(t)$ and $I_x(t)$ through the unitary transformation
$W(\theta,I_x)$ from the ground (non-rotating) state $\ket{\phi_0}$.
In the case of rotational motion,
$I_x$ corresponds to the angular momentum about the rotating axis $x$,
which is a conserved quantity, and $\theta$ is
the conjugate angle variable around the $x$-axis.
In order to guarantee the rotational invariance,
$W(\theta,I_x)$ has to be of the form
\beq
   W(\theta,I_x) = e^{-i\theta J_x} e^{iG(I_x)},
\label{eq:sccUform}
\eeq
where $J_x$ is the angular momentum operator about the $x$-axis,
and $G(I_x)$ is a one-body Hermite operator by which the intrinsic
state is specified:
\beq
   \ket{\phi(\theta,I_x)}=e^{-i\theta J_x}\ket{\phi_\intr(I_x)},
   \quad \ket{\phi_\intr(I_x)}=e^{iG(I_x)}\ket{\phi_0}.
\label{eq:sccWFb}
\eeq
The generators of the unitary transformation $W(\theta,I_x)$ are
defined by $(\partial W/\partial q) W^{-1}$ for $q=\theta$ or $I_x$,
and they have, from Eq.~(\ref{eq:sccUform}), the form
\ben
  \frac{\partial W}{\partial I_x} W^{-1}&=&
  e^{-i\theta J_x} \frac{\partial e^{iG(I_x)}}{\partial I_x}
  e^{-iG(I_x)} e^{i\theta J_x}
     \equiv i\Theta(I_x).
\label{eq:sccGena} \\
  i\frac{\partial W}{\partial \theta}W^{-1}&=& J_x,
\label{eq:sccGenb}
\een

     One of the basic equations of the SCC method is
the canonical variable conditions,\cite{MMSK80} \ 
which declare that the introduced collective variables
are canonical coordinate and momentum.
In the present case they are given as
\ben
  &&\braketc{\phi(\theta,I_x)}{i\Theta(I_x)}{\phi(\theta,I_x)}=0,
\label{eq:sccCVCa} \\
  &&\braketc{\phi(\theta,I_x)}{J_x}{\phi(\theta,I_x)}=I_x,
\label{eq:sccCVCb}
\een
and form which the week canonical variable condition is derived:
\beq
  \braketc{\phi(\theta,I_x)}{\,[J_x,i\Theta(I_x)]\,}{\phi(\theta,I_x)}=1.
\label{eq:sccWCVC}
\eeq
The other basic equations, the canonical equations of motion for
the collective variables and the equation of path, are derived
by the TDHB variational principle,\footnote{
 In this subsection the unit of $\hbar=1$ is used.
}
\beq
  \delta \braketc{\phi(\theta,I_x)}{\bigr( H - i\frac{d}{dt} \bigl)}
  {\phi(\theta,I_x)}=0,
\label{eq:varTDHB}
\eeq
or by using the generators, Eqs.~(\ref{eq:sccGena}) and~(\ref{eq:sccGenb}),
\beq
  \braketc{\phi(\theta,I_x)}
  {\,[O,\, H - {\dot \theta}J_x + {\dot I_x}\Theta(I_x)\,]\,}
  {\phi(\theta,I_x)}=0,
\label{eq:varTDHBo}
\eeq
where $O$ is an arbitrary one-body operator.
Taking the generators as $O$ and using the canonical variable conditions,
Eqs.~(\ref{eq:sccCVCa})$-$(\ref{eq:sccWCVC}),
one obtains the canonical equations of motion:
\ben
  {\dot \theta} &=& \frac{\partial \cH}{\partial I_x} =\omega_\rot(I_x),
\label{eq:eqmotiona} \\
  {\dot I_x} &=& -\frac{\partial \cH}{\partial \theta}=0,
\label{eq:eqmotionb} 
\een
with
\beq
  \cH(I_x) \equiv \braketc{\phi(\theta,I_x)}{H}{\phi(\theta,I_x)}
   =\braketc{\phi_\intr(I_x)}{H}{\phi_\intr(I_x)},
\label{eq:colH}
\eeq
where the rotational invariance of the Hamiltonian, $[H,J_x]=0$, is used.
Equation~(\ref{eq:eqmotionb}) is nothing else than the angular momentum
conservation, and Eq.~(\ref{eq:eqmotiona}) tells us that
the rotational frequency is constant, i.e. the uniform rotation.
Making use of these equations of motion, the variational principle
reduces to the equation of path
\beq
  \delta \braketc{\phi_\intr(I_x)} {\,H - \omega_\rot(I_x) J_x\,}
  {\phi_\intr(I_x)}=0,
\label{eq:eqpath}
\eeq
namely it leads precisely to the cranking model.
The remaining task is to solve this equation to obtain
the operator $iG(I_x)$ under the canonical variable conditions,
which are now rewritten as
\ben
  &&\braketc{\phi_\intr(I_x)}{C(I_x)}{\phi_\intr(I_x)}=0,
  \quad C(I_x)\equiv \frac{\partial e^{iG(I_x)}}{\partial I_x} e^{-iG(I_x)},
\label{eq:sccrCVCa} \\
  &&\braketc{\phi_\intr(I_x)}{J_x}{\phi_\intr(I_x)}=I_x.
\label{eq:sccrCVCb}
\een

   In Ref.~\citen{TMS91}, Eqs.~(\ref{eq:eqpath})$-$(\ref{eq:sccrCVCb})
are solved by means of the power series expansion method
with respect to $I_x$, which gives the functional form of
the rotational frequency $\omega_\rot(I_x)$.
It is, however, well known that the convergence radius of
the power series expansion with respect to $\omega_\rot$ is much larger,
so that the applicability of the method can be enlarged.\cite{BMtextII} \ 
Thus, the independent variable is changed to be $\omega_\rot$
instead of $I_x$ in the equations above.
In the following, we write the rotational frequency as $\omega$  
in place of $\omega_\rot$ for making the notation simpler.
Now the basic equations can be rewritten as
\ben
  &\delta& \braketc{\phi_\intr(\omega)} {H - \omega J_x}
  {\phi_\intr(\omega)}=0,
\label{eq:eqpathF} \\
  &&\braketc{\phi_\intr(\omega)}{C(\omega)}{\phi_\intr(I_x)}=0,
  \quad C(\omega)\equiv \frac{\partial e^{iG(\omega)}}{\partial \omega}
  e^{-iG(\omega)},
\label{eq:sccCVCaF} \\
  &&\braketc{\phi_\intr(\omega)}{J_x}{\phi_\intr(\omega)}=I_x(\omega).
\label{eq:sccCVCbF}
\een
Note that the last equation is not the constraint now, but
it just gives the functional form of the angular momentum $I_x$
in terms of $\omega$.
The first two equations, Eqs.~(\ref{eq:eqpathF}) and~(\ref{eq:sccCVCaF}),
are enough to get $iG(\omega)$, which makes the calculation simpler.
The equation of motion is transformed to the canonical relation
\beq
  \frac{\partial \cH'}{\partial \omega} = - I_x(\omega),
  \quad
\label{eq:sccCanRel}
\eeq
with the total Routhian in the rotating frame
\beq
  \cH'(\omega) \equiv
  \braketc{\phi_\intr(\omega)} {H - \omega J_x} {\phi_\intr(\omega)}.
\label{eq:colRouth}
\eeq
In order to show this, we note the following identity,
\beq
   \frac{\partial \braketc{\phi_\intr(\omega)}{O}{\phi_\intr(\omega)}}
  {\partial \omega}
  = \braketc{\phi_\intr(\omega)}{\,[O,\,C(\omega)]\,}{\phi_\intr(\omega)},
\label{eq:dervOmega}
\eeq
for an arbitrary $\omega$-independent one-body operator $O$. Then,
\beq
  \frac{\partial \cH'}{\partial \omega}
   =\braketc{\phi_\intr(\omega)}
   {\,[H - \omega J_x,\,C(\omega)]\,}{\phi_\intr(\omega)}
   -\braketc{\phi_\intr(\omega)}{J_x}{\phi_\intr(\omega)}
\label{eq:sccCanRelp}
\eeq
which lead to Eq.~(\ref{eq:sccCanRel}) because
the first term of the right hand side vanishes due to
the variational equation~(\ref{eq:eqpathF}).

    The one-body operator $iG(\omega)$ generates the unitary transformation
from the non-rotating (ground) state $\ket{\phi_0}$,
see Eq.~(\ref{eq:sccWFb}), and it is composed of
the $a^\dagger_i a^\dagger_j$ and $a_j a_i$ terms,
where $a^\dagger_i$ and $a_i$ are the creation and annihilation
operators of the quasiparticle state $i$
with respect to the ground state $\ket{\phi_0}$ as a vacuum state.
The solution of the basic equations is obtained in the form of
power series expansion
\beq
  iG(\omega)=\sum_{n=1}^{\infty} iG^{(n)}(\omega),
\label{eq:expfG}
\eeq
with
\beq
  iG^{(n)}(\omega) = \omega^n \bigl\{ \sum_{i < j}
  g^{(n)}(ij)\,a^\dagger_i a^\dagger_j
  - {\rm h.c.} \bigr\}.
\label{eq:sccGn}
\eeq
It is convenient to
introduce a notation for the transformed operator,
which is also expanded in power series of $\omega$,
\beq
  \maru{O} (\omega) \equiv e^{-iG(\omega)}\,O\, e^{iG(\omega)}
  \equiv \sum_{n=0}^{\infty}
  \maru{O}{^{(n)}}(\omega),
\label{eq:expO}
\eeq
for which the following formula are useful;
\beq
  e^{-iG} O e^{iG} =
  \sum_{n=0}^{\infty} \frac{1}{n!}
     {\underbrace{[ \cdots [}_{n\rm\;times}}O,\,iG]\cdots iG],
\label{eq:eforma}
\eeq
and
\beq
  \maru{C}(\omega)=e^{-iG} \frac{\partial e^{iG}}{\partial \omega}
  = \sum_{n=0}^{\infty} \frac{1}{(n+1)!}
    {\underbrace{[ \cdots [}_{n\rm\;times}}
   \frac{\partial iG}{\partial \omega},\,iG]\cdots iG].
\label{eq:eformb}
\eeq
Then the basic equations for solving $iG(\omega)$ in the $n$-th order
in $\omega$ are
\ben
  &&\braketc{\phi_0}
  {\,[a_j a_i,\,\maru{H}{^{(n)}}
  - \omega \maru{J_x}{^{(n-1)}}]\,}
  {\phi_0}=0,
\label{eq:eqpathFg} \\
  &&\braketc{\phi_0}{\maru{C}{^{(n)}}}{\phi_0}=0,
\label{eq:sccCVCFg}
\een
and the canonical relation is
\beq
   \frac{\partial \cH^{\prime (n+1)}}{\partial \omega}
   = - I_x^{(n)}, \quad {\rm or}\quad
   (n+1)\cH^{\prime (n+1)} = - \omega I_x^{(n)},
\label{eq:sccCanRelo}
\eeq
where the total Routhian and the expectation value of the angular momentum
are also expanded in power series,
\beq
  \cH'(\omega)=\sum_{n=0}^{\infty} \cH^{\prime (n)}, \quad
  I_x(\omega)=\sum_{n=1}^{\infty} I_x^{(n)}. \quad
\eeq

   The lowest order solution is easily determined:
The $n=0$ and 1 parts of Eq.~(\ref{eq:sccCVCFg}) are satisfied trivially,
while the $n=1$ part of Eq.~(\ref{eq:eqpathFg}) is written as
\beq
  \braketc{\phi_0} {\,[a_j a_i,\,[H,\,iG^{(1)}]\,]\,}
  {\phi_0} = \omega
  \braketc{\phi_0}{\,[a_j a_i,\,J_x]\,}{\phi_0},
\label{eq:scc1st}
\eeq
or
\beq
  [H,\,iG^{(1)}]_\RPA = \omega J_{x\RPA},
\label{eq:scc1sta}
\eeq
where the subscript $[\;\;]_\RPA$ means that only the RPA order term is
retained; e.g. $J_{x\RPA}$ = $a^\dagger_i a^\dagger_j$
and $a_j a_i$ parts of $J_x$.
This is the RPA equation,\cite{RStext} \ 
with respect to the ground state $\ket{\phi_0}$,
for the angle operator $i\Theta_\RPA$
conjugate to the symmetry conserving mode $J_{x\RPA}$, and we obtain
\beq
  iG^{(1)} = \omega \cJ_0 \,i\Theta_\RPA, \quad I_x^{(1)}=\omega \cJ_0,
\label{eq:sol1st}
\eeq
where $\cJ_0$ is the Thouless-Valatin moment of inertia.
Note that the general solution of Eq.~(\ref{eq:scc1st}) contains
a term $i\omega c_J J_{x\RPA}$ with $c_J$ being an arbitrary real constant.
We have chosen $c_J=0$ as a physical boundary condition,
because $J_x$ operator generates the transformation from the intrinsic to
the laboratory frame and should be eliminated from the unitary transformation
generating the intrinsic state, see Eq.~(\ref{eq:sccWFb}).
Once the lowest order solution ($n=1$) is obtained, higher order solutions
($n \ge 2$) can be uniquely determined
by rewriting Eqs.~(\ref{eq:eqpathFg}) and~(\ref{eq:sccCVCFg})
in the following forms;
\ben
  &&\braketc{\phi_0} {\,[a_j a_i,\,[H,\,iG^{(n)}]\,]\,} {\phi_0}
  = \braketc{\phi_0}{\,[a_j a_i,B^{(n)}]\,}{\phi_0},
\label{eq:eqpathFf} \\
  &&\braketc{\phi_0}{\,[iG^{(n)},\,i\Theta_\RPA]\,}{\phi_0}
  =\frac{1}{(n-1)\cJ_0}\braketc{\phi_0}{D^{(n)}}{\phi_0},
\label{eq:sccCVCFf}
\een
with
\ben
 && B^{(n)} \equiv \maru{H}{^{(n)}}
   - [H,\,iG^{(n)}] - \omega \maru{J_x}{^{(n-1)}},
\label{eq:defB} \\
 && D^{(n)} \equiv \maru{C}{^{(n)}}
   - \bigl[\frac{\partial iG^{(n)}}{\partial \omega},\,iG^{(1)} \bigr]
   - \bigl[\frac{\partial iG^{(1)}}{\partial \omega},\,iG^{(n)} \bigr].
\label{eq:defD}
\een
Here $B^{(n)}$ and $D^{(n)}$ only contain $iG^{(m)}$ with $m \le n-1$,
and $\partial iG^{(n)}/\partial \omega = n\,iG^{(n)}/\omega$
and Eq.~(\ref{eq:sol1st}) are used.
Equation~(\ref{eq:eqpathFf}) has the same structure as
Eq.~(\ref{eq:scc1st}) or~(\ref{eq:scc1sta}) and is
an inhomogeneous linear equation for the amplitude $g^{(n)}(ij)$,
where the inhomogeneous term is determined by the lower order solutions
(see \S\ref{sec:SCCsol} for details).

As in the case of the first order equation,
if $iG^{(n)}$ is expanded in terms of the complete set of
the RPA eigenmodes which is composed of the non-zero normal modes
and the zero mode ($J_{x\RPA}$, $i\Theta_\RPA$),
the general solution of $iG^{(n)}$ contains the term
proportional to $J_{x\RPA}$,
and it is determined by Eq.~(\ref{eq:sccCVCFf}).
Once the boundary condition for $iG^{(1)}$ is chosen as above, however,
the term proportional to $J_{x\RPA}$ should vanish.
In order to show this, one has to note that
matrix elements of the Hamiltonian and of the angular momentum
can be chosen to be real
with respect to the quasiparticle basis ($a^\dagger_i$, $a_i$)
in a suitable phase convention, e.g. that of Ref.~\citen{BMtextI}.
Then the matrix elements of the RPA normal mode operators and
the angle operator $i\Theta_\RPA$ are also real, and so does
the matrix elements of $iG^{(1)}$.
If $iG^{(n)}$ is expanded in terms of the RPA eigenmodes,
the imaginary part of its matrix elements arises only from
the term proportional to $J_{x\RPA}$ because $iG^{(n)}$ is anti-Hermite
while $J_{x\RPA}$ is Hermite.
If we assume that $iG^{(m)}$ with $m \le n-1$ has no $J_{x\RPA}$
term so that its matrix elements are real,
then the right hand side of Eq.~(\ref{eq:sccCVCFf}) vanishes,
because $D^{(n)}$ is an anti-Hermite operator with real matrix elements
composed of $iG^{(m)}$ with $m \le n-1$.
Therefore, $iG^{(n)}$ neither contains
the $J_{x\RPA}$ term.  Thus, the fact that the operator $iG$ has no 
$J_{x\RPA}$ term is proved by induction.
The situation is exactly the same for the case of gauge rotation;
the $N_\RPA$ term ($N$ is either the neutron or proton number operator)
also does not appear in $iG$.
The method to solve the above basic equations for our case of
the separable interaction~(\ref{eq:resint}) will be discussed
in detail in \S\ref{sec:SCCsol}.

\subsection{ Diabatic quasiparticle states in the rotating frame}
\label{sec:SCCqpe}

     In the previous subsection the rotational motion based on
the ground state $\ket{\phi_0}$ is considered in terms of the SCC method.
The same treatment can be done for one-quasiparticle states.
The one-quasiparticle state is written in the most general form as 
\beq
   \ket{\phi_\oneqp(\omega)} =
   e^{iG(\omega)} \sum_i f_i(\omega) a^\dagger_i \,\ket{\phi_0},
\label{eq:sccOneQPWF}
\eeq
where $iG(\omega)$ as well as the amplitudes $f_i(\omega)$ are
determined by the TDHB variational principle.
Generally $iG(\omega)$ for the one-quasiparicle state is not the same
as that of the ground state rotational band because of the blocking effect.
However, we neglect this effect and use the same $iG(\omega)$
in the present work following the idea of
the independent quasiparticle motion in the rotating frame.\cite{BF79} \ 
Then by taking the variation
\beq
  \delta \left[
  \frac{\braketc{\phi_\oneqp(\omega)}{H-\omega J_x}{\phi_\oneqp(\omega)}}
     {\braketb{\phi_\oneqp(\omega)}{\phi_\oneqp(\omega)}} \right] =0
\eeq
with respect to the amplitudes $f_i$, one obtains an eigenvalue equation,
\beq
  \sum_j \epsilon'_{ij}(\omega) f_{j\mu}(\omega)
  = f_{i\mu}(\omega)E'_\mu(\omega),
\label{eq:eigQPE}
\eeq
with
\beq
   \epsilon'_{ij}(\omega) =
   \braketc{\phi_0}
   {a_i \bigl(\maru{H}(\omega) - \omega \maru{J_x}(\omega)\bigr) a^\dagger_j}
   {\phi_0}.
\label{eq:QPemat}
\eeq
Namely the excitation energy $E'_\mu(\omega)$
and the amplitudes $f_{i\mu}(\omega)$
of the rotating quasiparticle state $\mu$ are obtained by diagonalizing
the cranked quasiparticle Hamiltonian defined by
\ben
  \maru{h}{^\prime}(\omega) &\equiv& {\rm one\mbox{-}body\;part\;of\;}
   [e^{-iG(\omega)}(H - \omega J_x)\,e^{iG(\omega)} ]
  \nonumber \\
   &=& \sum_{ij} \epsilon'_{ij}(\omega)\,a^\dagger_i a_j,
\label{eq:defQPh}
\een
where, due to the equation of path,
Eq.~(\ref{eq:eqpathF}) or~(\ref{eq:eqpathFg}),
${\displaystyle \maru{h}{^\prime}(\omega)}$ has
no $a^\dagger a^\dagger$ and $aa$ terms.
Introducing the quasiparticle operator in the rotating frame,
\beq
   \alpha^\dagger_\mu(\omega)=
   e^{iG(\omega)} \sum_i f_{i\mu}(\omega) a^\dagger_i \,e^{-iG(\omega)},
\label{eq:rotQPop}
\eeq
we can see that the one-quasiparticle state~(\ref{eq:sccOneQPWF}) is
written as
\beq
  \ket{\phi_\oneqp(\omega)}
  = \alpha^\dagger_\mu(\omega)\ket{\phi_\intr(\omega)},
\eeq
and
\ben
  h'(\omega) &\equiv& {\rm one\mbox{-}body\;part\;of\;}(H - \omega J_x)
      \nonumber \\
  &=& \sum_{\mu} E'_\mu(\omega)\,
  \alpha^\dagger_\mu(\omega) \alpha_\mu(\omega).
\label{eq:rotQPh}
\een
Namely, the quasiparticle states in the rotating frame are
nothing but those given in the selfconsistent cranking model.
Thus, if $H$ contains residual interactions,
the effects of change of the mean-field are automatically included
in the quasiparticle Routhian operator~(\ref{eq:defQPh})
in contrast to the simple cranked shell model
where the mean-field parameters are fixed at $\omega=0$.

    It is crucially important to notice that the cutoff of the power
series expansion in evaluating Eq.~(\ref{eq:defQPh}) results in
the diabatic quasiparticle states; i.e. the positive and negative
quasiparticle solutions do not interact with each other
as functions of the rotational frequency.
This surprising fact has been found in Ref.~\citen{SM85}
and utilized in subsequent various applications to the problem
of high-spin spectroscopy; see e.g. Ref.~\citen{MSM88}.
Thus, we use
\beq
  [\maru{h}{^\prime}(\omega)]^{(n \le \nmax)}
  = \sum_{ij} \bigl(\sum_{n=0}^{\nmax}
  \omega^n \epsilon^{\prime (n)}_{ij}\bigr)a^\dagger_i a_j,
\label{eq:DiabQPh}
\eeq
with
\beq
   \epsilon^{\prime (n)}_{ij} \equiv
   \braketc{\phi_0}
   {a_i \bigl(\maru{H}{^{(n)}} - \omega \maru{J_x}{^{(n-1)}}\bigr)
   a^\dagger_j} {\phi_0}/\omega^n,
\label{eq:DiabQPemat}
\eeq
as a diabatic quasiparticle Routhian operator.
If we take $\nmax=1$ and use the solution~(\ref{eq:sol1st}),
the first order Routhian operator is explicitly written as
\beq
  [\maru{h}{^\prime}(\omega)]^{(n \le 1)}
  = h - \omega (J_x - J_{x\RPA}),
\label{eq:DiabQPh1st}
\eeq
with $h \equiv {\rm one\mbox{-}body\;part\;of\;}H$.
This Hamiltonian was used to construct a diabatic quasiparticle
basis in Ref.~\citen{TS81} to study the $g$-$s$ band crossing problem.
We will show in \S\ref{sec:SCCappli} that the inclusion of higher order
terms improves the quasiparticle Routhian in comparison with
experimental data.

    In order to study properties of one-body observables
in the rotating frame, for example, the aligned angular momenta
of quasiparticles,
an arbitrary one-body operator $O$ has to be
expressed in terms of the diabatic quasiparticle basis~(\ref{eq:rotQPop});
\ben
  O &=& e^{iG(\omega)} \maru{O}(\omega) e^{-iG(\omega)}
    \nonumber \\
   &=& \braketc{\phi_\intr(\omega)}{O}{\phi_\intr(\omega)}
   + \sum_{\mu\nu}O_B(\mu\nu;\omega) \,\alpha^\dagger_\mu \alpha_\nu
    \nonumber \\
   &+& \sum_{\mu<\nu} \bigl(
      O_{A+}(\mu\nu;\omega) \,\alpha^\dagger_\mu \alpha^\dagger_\nu
    + O_{A-}(\mu\nu;\omega) \,\alpha_\nu \alpha_\mu \bigr)
\label{eq:OrotRep}
\een
where the matrix elements are written as
\ben
  && O_{B}(\mu\nu;\omega)
   =\sum_{ij} f^*_{i\mu}(\omega) f_{j\nu}(\omega)
   \braketc{\phi_0}{\,a_i
  \bigl(\sum_{n=0}^{\nmax} \maru{O}{^{(n)}}(\omega)\bigr)
   a^\dagger_j\,}{\phi_0},
\label{eq:OrotB} \\
  && O_{A+}(\mu\nu;\omega)
   =\sum_{ij} f^*_{i\mu}(\omega) f^*_{j\nu}(\omega)
   \braketc{\phi_0}{\,[a_j a_i,\,
  \bigl(\sum_{n=0}^{\nmax} \maru{O}{^{(n)}}(\omega)\bigr)
    ]\,}{\phi_0},
\label{eq:OrotA+} \\
  && O_{A-}(\mu\nu;\omega)
   =\sum_{ij} f_{i\mu}(\omega) f_{j\nu}(\omega)
   \braketc{\phi_0}{\,
  \bigl(\sum_{n=0}^{\nmax} \maru{O}{^{(n)}}(\omega)\bigr)
   ,\,a^\dagger_i a^\dagger_j]\,}{\phi_0}.
\label{eq:OrotA-}
\een
It is clear from this expression that there are two origins
of the $\omega$-dependence of the matrix elements;
one is the effect of collective rotation, Eq.~(\ref{eq:expO}),
which is treated in the power series expansion in $\omega$ and truncated
up to $\nmax$, and the other comes from the diagonalization of
the quasiparticle Routhian operator, Eq.~(\ref{eq:eigQPE}).
Our method to calculate the rotating quasiparticle states can be
viewed as a two-step diagonalization; the first step is
the unitary transformation $e^{iG(\omega)}$,
which eliminates the dangerous terms,
the $a^\dagger a^\dagger$ and $a a$ terms,
of the Routhian operator ${\displaystyle \maru{h}{^{\prime}}}$
up to the order $\nmax$ in $\omega$ leading to the diabatic basis,
while the second step diagonalizes its one-body part,
the $a^\dagger a$ -terms.
We shall discuss this two-step transformation in more detail
in \S\ref{sec:constDQPbasis}.
In this way we can cleanly separate the effects of
the collective rotational motion on the intrinsic states of the $g$-band
and on the independent quasiparticle motion in the rotating frame.
As long as the one-step diagonalization is performed as in the case
of the usual cranking model, this separation cannot be achieved
and the problem of the unphysical interband mixing is inevitable.

\subsection{ Solution of the equation of path
             by means of the RPA response function}
\label{sec:SCCsol}

     Now we present a concrete procedure to solve the equation of path,
Eq.~(\ref{eq:eqpathF}), for our Hamiltonian which is composed of
the Nilsson single-particle potential and
the multi-component separable interaction~(\ref{eq:resint}).
Let us rewrite our total Hamiltonian in the following form:
\beq
  H = h - \frac{1}{2}\sum_\rho\chi_\rho Q_\rho Q_\rho,
\label{eq:sepHform}
\eeq
where $Q_\rho$ are Hermite operators satisfying
\beq
     Q_\rho = Q^\dagger_\rho, \quad \braketc{\phi_0}{Q_\rho}{\phi_0}=0,
\label{eq:sepQcond}
\eeq
and $\ket{\phi_0}$ is the HB ground state of $H$.\footnote{
  We employ the HB approximation, i.e. do not include the exchange terms
  of the separable interactions throughout this paper.  }
The mean-field Hamiltonian $h$ includes the pairing potential and
the number constraint term as well as the Nilsson Hamiltonians:
\beq
  h=h_\Nils -\sum_\tau\sum_{L=0,2} \Delta_{L0\tau}
  \bigl( P^{\tau\dagger}_{L0}+P^\tau_{L0} \bigr)
  - \sum_\tau\lambda_\tau N_\tau,
\label{eq:meanfh}
\eeq
where the nuclei under consideration are assumed to be
axially symmetric at $\omega=0$.
Our Hamiltonians has a symmetry with respect to the $180^\circ$-rotation around
the rotation-axis ($x$-axis), the quantum number of which is called
{\it signature}, $r=e^{-i\alpha}$;
therefore the operators $Q_\rho$ are classified according to
the signature quantum numbers,\cite{SM83} \ $r=\pm 1$ or $\alpha=0,1$.
Moreover, we can choose the phase convention\cite{BMtextI}
in such a way that the matrix elements of the Hamiltonians $H$
and of the angular momentum $J_x$ are real.
Then the operators $Q_\rho$ are further classified into two categories,
i.e. real and imaginary operators,
whose matrix elements are real and pure imaginary, respectively.
Since expectation values of
the signature $r=-1$ ($\alpha=1$) operators and of the imaginary operators
vanish in the cranking model, operators
with signature $r=+1$ and real matrix elements only contribute 
to the equation of path for the collective rotation.
This observation is important.  As shown in the end of
\S\ref{sec:SCCbasic}, the boundary condition~(\ref{eq:sol1st})
for the collective rotation
leads that the transformation operator $iG(\omega)$ does not
contain the $J_{x\RPA}$ part in all orders.
Absence of the imaginary operators guarantees that the matrix elements
of $iG(\omega)$ are real and 
Eq.~(\ref{eq:sccCVCaF}) is automatically satisfied: 
We need not use this equation anymore.

   Thus, the operators that are to be included in Eq.~(\ref{eq:sepHform})
in order to solve the basic equations for $iG(\omega)$ are
\beq
  \{Q_\rho\} = P^\tau_{00+},\; P^{(+)\tau}_{20+},\; P^{(+)\tau}_{21+},
  \; P^{(+)\tau}_{22+},\; Q^{(+)}_{20}, Q^{(+)}_{22},
\label{eq:intOps}
\eeq
and correspondingly the strengths are
\beq
  \{\chi_\rho\} = G^\tau_0/2,\; G^\tau_2/2,\; G^\tau_2/2,\; G^\tau_2/2,
  \; \kappa_{20},\; \kappa_{22},
\label{eq:intStrengths}
\eeq
where $\tau=\nu,\,\pi$ distinguishes the neutron and proton operators.
Here the following definitions are used;
for the pairing operators,
\ben
  P_{00+} = P^\dagger_{00} + P_{00}&,& \quad
  P_{00-} = i\bigl( P^\dagger_{00} - P_{00} \bigr),
    \nonumber \\
  P^{(\pm)}_{2K+} = P^{(\pm)\dagger}_{2K} + P^{(\pm)}_{2K}&,& \quad
  P^{(\pm)}_{2K-} = i\bigl( P^{(\pm)\dagger}_{2K} - P^{(\pm)}_{2K} \bigr),
\label{eq:sepPpm}
\een
and for signature coupled operators,
\ben
  P^{(\pm)}_{2K} &=& \frac{1}{\sqrt{1+\delta_{K0}}}
  \bigl( \colon P_{2K}\colon  \pm \colon P_{2-K}\colon \bigr), \quad (K \ge 0)
    \nonumber \\
  Q^{(\pm)}_{2K} &=& \frac{1}{\sqrt{1+\delta_{K0}}}
  \bigl( \colon Q_{2K}\colon  \pm \colon Q_{2-K}\colon \bigr), \quad (K \ge 0)
\label{eq:sepQsig}
\een
in which the superscript $(\pm)$ denotes the signature $r=\pm 1$,
and $\colon O \colon \equiv O - \braketc{\phi_0}{O}{\phi_0}$.
The quasiparticle creation and annihilation operators should also be
classified according to the signature quantum number;
$a^\dagger_i$ for $r=+i$ ($\alpha=-1/2$) and
$a^\dagger_{\bar i}$ for $r=-i$ ($\alpha=+1/2$).
Then the mean-field Hamiltonian $h$
is expressed in terms of them as
\beq
  h = \sum_{i>0} \bigl( E_i a^\dagger_i a_i
  + E_{\bar i} a^\dagger_{\bar i} a_{\bar i} \bigr),
\label{eq:Ophform}
\eeq
where $\sum_{i>0}$ means that only half of the single-particle levels
has to be summed corresponding to the signature classification,
and the quasiparticle energy at $\omega=0$ satisfies $E_i=E_{\bar i}$.
In the same way, $Q_\rho$ are written as
\ben
  Q_\rho = \sum_{ij>0} q^A_\rho(ij)\,
  (a^\dagger_i a^\dagger_{\bar j} + a_{\bar j} a_i)
  +\sum_{ij>0}\bigl(q^B_\rho(ij)\, a^\dagger_i a_j +
    {\bar q}^B_\rho(ij)\, a^\dagger_{\bar i} a_{\bar j} \bigr),
\label{eq:OpQform}
\een
where the matrix elements satisfy, at $\omega=0$,
$q^A_\rho(ji)=\pm q^A_\rho(ij)$ and
${\overline q}^B_\rho(ij) =\pm \, q^B_\rho(ij)$
for $Q_\rho$ with the time-reversal property being $\pm$,
if the phase convention of Ref.~\citen{BMtextI} is used.

   Now let us consider the method to solve the equations for $iG(\omega)$.
As is already discussed in \S\ref{sec:SCCbasic},
the solution is sought in the form of power series expansion in $\omega$,
where the $n$-th order term $iG^{(n)}$ is written as
\ben
  iG^{(n)}= \omega^n \sum_{ij>0} g^{(n)}(ij)\,
  (a^\dagger_i a^\dagger_{\bar j} - a_{\bar j} a_i).
\label{eq:OpGnform}
\een
The $n$-th order equation Eq.~(\ref{eq:eqpathFf}) has the structure
of an inhomogeneous linear equation for the amplitudes $g^{(n)}(ij)$,
\beq
{\mib K} \left( \bea{r} g^{(n)} \\ -g^{(n)} \eea \right)
  = \left( \bea{r} b^{(n)} \\ -b^{(n)} \eea \right),
\label{eq:sccnth}
\eeq
where ${\mib K}$ is the RPA energy matrix
\ben
&{\mib K}&(ij;kl) = \left(
\bea{cc} A(ij;kl)   & B(ij;kl)   \\
         B^*(ij;kl) & A^*(ij;kl) \eea
 \right) \nonumber \\
 &=& \left(
\bea{cc}
  \braketc{\phi_0}{\,[a_{\bar j} a_i,\,
     [H,\,a^\dagger_k a^\dagger_{\bar l}]\,]\,}{\phi_0 } &
  \braketc{\phi_0}{\,[a_{\bar j} a_i,\,
     [H,\,a_{\bar l} a_k]\,]\,}{\phi_0} \\
  \braketc{\phi_0}{\,[a^\dagger_i a^\dagger_{\bar j},\,
     [H,\,a^\dagger_k a^\dagger_{\bar l}]\,]\,}{\phi_0} &
  \braketc{\phi_0}{\,[a^\dagger_i a^\dagger_{\bar j},\,
     [H,\,a_{\bar l} a_k ]\,]\,}{\phi_0} \eea \right),
\label{eq:RPAham}
\een
and the amplitudes $b^{(n)}(ij)$ in the inhomogeneous term are defined by
\beq
  a^\dagger a^\dagger \;{\rm and}\; a a\, {\rm \mbox{}parts\; of\;}
  B^{(n)}= \omega^n \sum_{ij>0} b^{(n)}(ij)\,
  (a^\dagger_i a^\dagger_{\bar j} + a_{\bar j} a_i).
\label{eq:OpBnform}
\eeq
For the first order $n=1$, $B^{(1)}=\omega J_{x\RPA}$
and Eq.~(\ref{eq:sccnth}) determines the RPA angle operator $i\Theta_\RPA$,
as discussed in \S\ref{sec:SCCbasic}.
Since the part of interaction
composed of the imaginary operators,
e.g. $P_{00-}$, $P_{20-}$ and $Q^{(+)}_{21}$ etc.,
which are related to the symmetry recovering mode $J_{x\RPA}$ (and $N_\RPA$)
are not included, the RPA matrix ${\mib K}$ (with signature $r=+1$)
has no zero-modes and can be inverted without any problem.
However, the dimension of the RPA matrix is not small in realistic
situations, and therefore we invoke the merit of separable interactions;
by using the response-function matrix for the $Q_\rho$ operators,
the inversion of the RPA matrix is reduced to the inversion of
the response-function matrix itself whose dimension is much smaller.
Inserting the Hamiltonian~(\ref{eq:sepHform}) into Eq.~(\ref{eq:eqpathFf}),
we obtain
\beq
 (E_i + E_{\bar j})g^{(n)}(ij)
   - \sum_\rho q^A_\rho(ij) \chi_\rho \cQ^{(n)}_\rho = b^{(n)}(ij),
\label{eq:sepgnth}
\eeq
where
\beq
  \cQ^{(n)}_\rho \equiv
  \braketc{\phi_0}{\,[Q_\rho,\,iG^{(n)}]\,}{\phi_0}/\omega^n
  =2 \sum_{ij>0}q^A_\rho(ij)g^{(n)}(ij).
\label{eq:calQnth}
\eeq
Then inhomogeneous linear equations for $\cQ^{(n)}_\rho$
can be easily derived as
\beq
  \sum_\sigma
  (\delta_{\rho\sigma} - R_{\rho\sigma}\chi_\sigma)\,\cQ^{(n)}_\sigma
  = \cB^{(n)}_\rho,
\label{eq:eqQnth}
\eeq
where
\beq
  R_{\rho\sigma} \equiv 2 \sum_{ij>0}
  \frac{q^A_\rho(ij)q^A_\sigma(ij)}{E_i+E_{\bar j}},\quad
  \cB^{(n)}_\rho \equiv 2 \sum_{ij>0}
  \frac{b^{(n)}(ij)q^A_\rho(ij)}{E_i+E_{\bar j}}.
\label{eq:respRB}
\eeq
Note that $R_{\rho\sigma}$ are the response functions for
operators $Q_\rho$ and $Q_\sigma$ at zero excitation energy,
and nothing but the inverse energy weighted sum rule values (polarizability).
Equation~(\ref{eq:eqQnth}) is much more easily solved than
Eq.~(\ref{eq:sccnth}) because of the huge reduction of dimension,
and we obtain
\beq
 g^{(n)}(ij)=\frac{1}{E_i+E_{\bar j}} \bigl\{ \sum_{\rho\sigma}
  q^A_\rho(ij)\chi_\rho[\,(1-R\chi)^{-1}]_{\rho\sigma}\cB^{(n)}_\sigma
  + b^{(n)}(ij) \bigl\},
\label{eq:solgnth}
\eeq
where the matrix notations are used for $R=(R_{\rho\sigma})$
and $\chi = (\delta_{\rho\sigma}\chi_\rho)$.
Apparently the $n=1$ solution gives the Thouless-Valtin moment of inertia,
\beq
   \cJ_0 =\cJ_{\rm TV}=\cJ_{\rm Bely} + \cJ_{\rm Mig}, \quad
   \cJ_{\rm Bely} =  2 \sum_{ij>0}
  \frac{J_x^A(ij)J_x^A(ij)}{E_i+E_{\bar j}},
\label{eq:THVmom}
\eeq
and
\beq
  \cJ_{\rm Mig} =  \sum_{\rho\sigma}
  \cB^J_\rho\,\chi_\rho[\,(1-R\chi)^{-1}]_{\rho\sigma}\cB^J_\sigma,
  \quad {\rm with}\;\;
  \cB^J_\rho \equiv 2 \sum_{ij>0}
  \frac{J_x^A(ij)q^A_\rho(ij)}{E_i+E_{\bar j}}.
\label{eq:THVmoma}
\eeq
where $J_x^A(ij)$ denote the $a^\dagger a^\dagger$ and $aa$ parts of $J_x$,
and the summation ($\rho,\sigma$) in Eq.~(\ref{eq:THVmoma})
runs, at $\omega=0$, only over $\rho,\sigma=P^{(+)\tau}_{21+}$,
namely the $K=1$ quadrupole-pairing component.
Once the perturbative solution of $iG(\omega)$ is obtained,
the quasiparticle energy can be calculated by diagonalizing
\beq
  \maru{h}{^\prime}(\omega)=
   \sum_{ij>0} \bigl( \epsilon'_{ij}\,(\omega)a^\dagger_i a_j
   +  {\bar \epsilon}'_{ij}\,(\omega)a^\dagger_{\bar i} a_{\bar j}
   \bigr),
\label{eq:QPEh}
\eeq
and one obtains
\beq
  h'(\omega)=
   \sum_{\mu>0} \bigl( E'_\mu(\omega)\,
    \alpha^\dagger_\mu(\omega) \alpha_\mu(\omega)
   +E'_{\bar \mu}(\omega)\,
    \alpha^\dagger_{\bar \mu}(\omega) \alpha_{\bar \mu}(\omega)
   \bigr),
\label{eq:rotQPE}
\eeq
where the first and second terms in these two equations
correspond to the quasiparticle states
with signature $r=+i$ $(\alpha=-1/2)$ and $r=-i$ $(\alpha=+1/2)$, 
respectively.

    At the end of this subsection a few remarks are in order: 
First, although it is assumed that the starting state $\ket{\phi_0}$
is the ground state at $\omega=0$, the formulation developed above 
can be equally well applied
also when the finite frequency state at $\omega=\omega_0$
is used as a starting state;
i.e. $\ket{\phi_0}$ is determined by
$\delta\braketc{\phi_0}{H-\omega_0 J_x}{\phi_0}$.
In such a case, however, the power series expansion should be performed
with respect to $(\omega - \omega_0)$.
In fact, the method has been applied in Ref.~\citen{Tera92}
to describe the $s$-band
by taking the starting state as the lowest two quasineutron state
at finite frequency, although the angular momentum
expansion in $(I-I_0)$ is used in it.
Secondly, as can be inferred from the form of the $n$-th order
solution~(\ref{eq:solgnth}), the $\omega$-expansion is based on
the perturbation with respect to the quantity $\omega/(E_i+E_j)$
(or $\omega/\omega_\lambda(\RPA)$, if the equation is solved
in terms of the RPA eigenmodes).
Therefore, it is expected that the convergence of the $\omega$-expansion
becomes poor when the average value of the two quasiparticle
energies is reduced: It is the case for the situation of weak pairing,
or when one takes the starting state at a finite frequency
where highly alignable two quasiparticle states have considerably
smaller excitation energies.
The difficulty in the calculation of $s$-band in Ref.~\citen{Tera92}
is possibly caused by this problem.
Thirdly, as mentioned already, the expectation value
of the nucleon number is not conserved along the rotational band.
This is because the number operator $N_\tau$ does not commute with
$iG(\omega)$; namely, there exists a coupling between the spatial
and the pairing rotations.  In order to achieve rigorous conservation
of nucleon numbers, one has to apply the SCC method also to
the pairing rotational motion,\cite{Mat86} \ 
and combine it to the present formalism.  In view of such a more general
formulation, the energy in the rotating frame~(\ref{eq:colRouth})
calculated in the present method is actually the double Routhian
$\cH''(\omega,\lambda_\tau=\lambda_{0\tau})$,
where $\lambda_{0\tau}$ is the chemical potential fixed
to conserve the number at the ground state $\omega=0$.
The $\omega$-dependence of the expectation value of number operator 
starts from the second order, and its coefficient is very small
as will be shown in \S\ref{sec:SCCappli}.
Therefore the effect of number non-conservation along the rotational band
is very small; this fact has been checked in Ref.~\citen{TMS91} 
by explicitly including the coupling to the pairing rotation. 
Finally, this method utilizing the response-function matrix
can be similarly applied to the case of the $(\eta^*,\eta)$-expansion
of the SCC method for problems of collective vibration.
In such a case, a full RPA response matrix (containing both real
and imaginary operators) is necessary, and
one has to choose one of the RPA eigenenergies,
to which the solution is continued in the small amplitude limit,
as the excitation energy of the response function.

\subsection{ Application to the $g$- and $s$- bands in rare-earth nuclei}
\label{sec:SCCappli}

     We apply the formulation of the SCC method for the collective rotation
developed in the previous subsections to even-even
deformed nuclei in the rare-earth region.
In this calculation, the same Nilsson potential
(the $ls$ and $ll$ parameters from Ref.~\citen{BR85})
is used as in \S\ref{sec:QPairInt},
but the hexadecupole deformation is not included.
As investigated in Ref.~\citen{TMS91}, the couplings of
collective rotation to the pairing vibrations as well as
the collective surface vibrations are important.
Therefore the model space composed of three oscillator shells,
$N_\osc=4-6$ for neutrons and $N_\osc=3-5$ for protons,
are employed and all the $\Delta N_\osc = 0,\pm 2$ matrix elements
of the quadrupole operators are included in the calculation.
In order to describe the properties of deformed nuclei,
the deformation parameter is one of the most important factors.
The Nilsson-Strutinsky calculation in \S\ref{sec:QPairInt}
gives slightly smaller values compared with the experimental data
deduced from the measured $B(E2,\,2^+_g \rightarrow 0^+_g)$
values.  Therefore, we take the experimental values for the $\epsilon_2$
parameter from Ref.~\citen{Loeb70}.
There exist, however, some cases where no experimental data are available.
Then we take the value obtained by extrapolation from available data
according to the scaling of the result of our Nilsson-Strutinsky calculation
in \S\protect\ref{sec:QPairInt};
for example, $\epsilon_2(^{154}$Dy) used is
$ \epsilon_2(^{154}$Dy$)^\rcal \times
\epsilon_2(^{156}$Dy$)^\rexp / \epsilon_2(^{156}$Dy$)^\rcal $.
The values adopted in the calculation are listed in Table~\ref{tab:SCCsum}.

%%%%%%%%%%%
%  Table  %
%%%%%%%%%%%

\begin{table}
\caption{
  Summary of the calculated results and comparison with experimental data
  for nuclei in the rare-earth region, Gd ($Z=64$) to W ($Z=74$).
  The deformation parameters $\epsilon_2$
  are taken from Ref.~\protect\citen{Loeb70}; superscript * denotes
  cases where no data is available and extrapolation based on
  our calculation in \S\protect\ref{sec:QPairInt} is employed.
  The Harris parameters $\cJ_0$ and $\cJ_1$ are given
  in unit of $\hbar^2/$MeV and $\hbar^4/$MeV$^3$, respectively.
  The energy gaps $\Delta$ are in unit of MeV.
  The third order even-odd mass differences
  based of the mass table of Ref.~\protect\citen{AW93} are used as
  experimental pairing gaps.
  }
\label{tab:SCCsum}
\begin{center}
\begin{tabular}{cccccccccccc}
\hline\hline
 & $N$  & $\epsilon_2$
 & ${\cal J}_0^{\rm cal}$
 & ${\cal J}_1^{\rm cal}$
 & ${\cal J}_0^{\rm exp}$
 & ${\cal J}_1^{\rm exp}$
 & ${\mit \Delta}_\nu^{\rm cal}$
 & ${\mit \Delta}_\pi^{\rm cal}$
 & ${\mit \Delta}_\nu^{\rm exp}$
 & ${\mit \Delta}_\pi^{\rm exp}$ \\
\hline
Gd & 88 & 0.164 & 11.8 & 308 &  8.7 & $-$ & 1.157 & 1.424 & 1.108 & 1.475 \\
   & 90 & 0.251 & 25.6 & 341 & 23.1 & 333 & 1.270 & 1.169 & 1.277 & 1.133 \\
   & 92 & 0.274 & 31.5 & 165 & 33.4 & 179 & 1.222 & 1.097 & 1.070 & 0.960 \\
   & 94 & 0.282 & 34.2 & 118 & 37.6 & 111 & 1.152 & 1.060 & 0.892 & 0.878 \\
   & 96 & 0.287 & 36.0 &  98 & 39.7 & 101 & 1.073 & 1.030 & 0.831 & 0.871 \\
Dy & 88 & 0.205$^*$& 17.4 & 134 & 9.0 & $-$ & 1.187 & 1.261 & 1.177 & 1.472 \\
   & 90 & 0.242 & 24.3 & 223 & 20.1 & 348 & 1.233 & 1.138 & 1.269 & 1.162 \\
   & 92 & 0.261 & 29.4 & 178 & 29.9 & 184 & 1.196 & 1.073 & 1.077 & 1.033 \\
   & 94 & 0.271 & 32.7 & 136 & 34.3 & 123 & 1.128 & 1.033 & 0.967 & 0.978 \\
   & 96 & 0.270 & 34.3 & 120 & 37.0 &  93 & 1.050 & 1.013 & 0.917 & 0.930 \\
   & 98 & 0.275 & 36.8 & 117 & 40.7 &  98 & 0.970 & 0.984 & 0.832 & 0.875 \\
Er & 88 & 0.162$^*$& 12.2 & 110 & 8.7 & $-$ & 1.105 & 1.321 & 1.213 & 1.396 \\
   & 90 & 0.204 & 18.6 & 112 & 13.0 & 281 & 1.153 & 1.188 & 1.277 & 1.244 \\
   & 92 & 0.245 & 26.2 & 154 & 23.1 & 196 & 1.165 & 1.075 & 1.138 & 1.137 \\
   & 94 & 0.258 & 30.3 & 130 & 29.0 & 133 & 1.105 & 1.031 & 1.078 & 1.091 \\
   & 96 & 0.269 & 33.5 & 104 & 32.6 &  93 & 1.028 & 0.995 & 1.035 & 0.987 \\
   & 98 & 0.272 & 35.8 & 108 & 37.1 & 105 & 0.951 & 0.971 & 0.966 & 0.877 \\
   &100 & 0.271 & 36.4 & 103 & 37.5 &  57 & 0.919 & 0.953 & 0.776 & 0.857 \\
   &102 & 0.268 & 35.4 &  76 & 38.1 &  59 & 0.907 & 0.938 & 0.708 & 0.797 \\
\hline
\end{tabular}
\end{center}
\end{table}

\addtocounter{table}{-1}
\begin{table}
\caption{
  {\it continued.}
  }
\begin{center}
\begin{tabular}{cccccccccccc}
\hline\hline
 & $N$  & $\epsilon_2$
 & ${\cal J}_0^{\rm cal}$
 & ${\cal J}_1^{\rm cal}$
 & ${\cal J}_0^{\rm exp}$
 & ${\cal J}_1^{\rm exp}$
 & ${\mit \Delta}_\nu^{\rm cal}$
 & ${\mit \Delta}_\pi^{\rm cal}$
 & ${\mit \Delta}_\nu^{\rm exp}$
 & ${\mit \Delta}_\pi^{\rm exp}$ \\
\hline
Yb & 90 & 0.172$^*$& 14.4 & 83 & 9.1 & 221 & 1.124 & 1.200 & 1.402 & 1.253 \\
   & 92 & 0.197$^*$& 18.9 & 119 & 16.6 & 204 & 1.136 & 1.128 & 1.168 & 1.180 \\
   & 94 & 0.218$^*$& 23.8 & 141 & 23.5 & 186 & 1.106 & 1.070 & 1.137 & 1.214 \\
   & 96 & 0.245$^*$& 30.1 & 119 & 29.0 & 131 & 1.024 & 1.012 & 1.159 & 1.111 \\
   & 98 & 0.258 & 33.6 & 111 & 34.0 & 127 & 0.950 & 0.981 & 1.039 & 0.983 \\
   &100 & 0.262 & 34.9 & 108 & 35.5 &  83 & 0.915 & 0.959 & 0.865 & 0.908 \\
   &102 & 0.267 & 34.5 &  75 & 38.0 &  70 & 0.889 & 0.938 & 0.764 & 0.840 \\
   &104 & 0.259 & 33.4 &  70 & 39.1 &  64 & 0.862 & 0.926 & 0.685 & 0.848 \\
   &106 & 0.250 & 32.3 &  93 & 36.4 &  55 & 0.847 & 0.918 & 0.585 & 0.815 \\
Hf & 92 & 0.163$^*$& 14.1 &  90 & 12.2 & 178 & 1.154 & 1.105 & 1.219 & 1.260 \\
   & 94 & 0.181$^*$& 17.8 & 129 & 17.7 & 196 & 1.148 & 1.057 & 1.175 & 1.285 \\
   & 96 & 0.207$^*$& 23.6 & 134 & 23.5 & 191 & 1.083 & 1.004 & 1.123 & 1.182 \\
   & 98 & 0.218$^*$& 27.0 & 122 & 29.3 & 194 & 1.032 & 0.976 & 1.022 & 1.062 \\
   &100 & 0.227 & 29.5 & 116 & 31.2 & 131 & 0.986 & 0.952 & 0.953 & 0.988 \\
   &102 & 0.235 & 30.6 &  94 & 32.7 & 110 & 0.935 & 0.932 & 0.901 & 0.915 \\
   &104 & 0.245 & 31.4 &  74 & 33.8 &  88 & 0.867 & 0.915 & 0.811 & 0.864 \\
   &106 & 0.227 & 29.2 &  99 & 32.1 &  65 & 0.867 & 0.903 & 0.693 & 0.824 \\
   &108 & 0.227 & 26.9 & 100 & 32.1 &  40 & 0.898 & 0.887 & 0.745 & 0.856 \\
W  & 92 & 0.148$^*$& 12.1 &  70 &  9.4 & 159 & 1.159 & 1.006 & 1.331 & 1.295 \\
   & 94 & 0.161$^*$& 14.6 & 100 & 13.2 & 182 & 1.169 & 0.968 & 1.201 & 1.142 \\
   & 96 & 0.179$^*$& 18.5 & 122 & 17.8 & 216 & 1.139 & 0.928 & 1.146 & 1.100 \\
   & 98 & 0.196$^*$& 22.6 & 118 & 23.4 & 255 & 1.082 & 0.899 & 1.046 & 1.053 \\
   &100 & 0.206$^*$& 25.4 & 110 & 26.3 & 171 & 1.032 & 0.880 & 1.091 & 1.023 \\
   &102 & 0.211$^*$& 26.7 &  99 & 27.1 & 134 & 0.985 & 0.865 & 0.931 & 1.027 \\
   &104 & 0.214$^*$& 27.3 &  83 & 28.0 & 112 & 0.929 & 0.850 & 0.884 & 1.036 \\
   &106 & 0.212 & 26.8 & 95 & 28.7 & 86 & 0.890 & 0.833 & 0.802 & 0.943 \\
   &108 & 0.208 & 24.5 & 92 & 29.8 & 53 & 0.903 & 0.817 & 0.814 & 0.849 \\
   &110 & 0.197 & 21.5 & 77 & 26.8 & 55 & 0.927 & 0.805 & 0.720 & 0.868 \\
   &112 & 0.191 & 19.6 & 76 & 24.3 & 67 & 0.919 & 0.794 & 0.793 & 0.907 \\
\hline
\end{tabular}
\end{center}

\end{table}

   The residual interaction is of the form given in Eq.~(\ref{eq:resint}),
where the double-stretched form factor is taken according to
the discussion in \S\ref{sec:QPairInt}.
However, we cannot use the same best values obtained
in \S\ref{sec:QPairInt} for the strengths of the pairing interactions,
since the model space and the treatment of $\Delta N_\osc = \pm 2$
matrix elements of the quadrupole operators are different.
Here we use $G^\nu_0 = 20/A$ MeV and $G^\pi_0 = 24/A$ MeV
for the monopole-pairing interaction,
by which monopole-pairing gaps calculated with the use of
the above model space roughly reproduce
the experimental even-odd mass differences 
(see Eq.~(\ref{eq:G0HO}), and note that an extra
difference of the constant ``$c$'' in it between neutrons and protons
comes from the difference of the model space).
As for the double-stretched quadrupole-pairing interaction,
we take $g^\nu_2 = g^\pi_2 = 24$ (see Eq.(\ref{eq:G2byG0})),
by which overall agreements are achieved for the moments of inertia.
The results are summarized in Table~\ref{tab:SCCsum}.
Here calculated energy gaps $\Delta$ are the monopole-pairing gaps,
but they are very similar to the average pairing gaps~(\ref{eq:avrGap})
because the double-stretched quadrupole-pairing interaction is used.
The isoscalar (double-stretched) quadrupole interaction
does not contribute to the Thouless-Valatin moment of inertia $\cJ_0$,
but affects the higher order Harris parameter $\cJ_1$.
We do not fit the strengths for each nucleus, but use
$\kappa_{2K}=1.45\,\kappa_2^\self$ (see Eq.~(\ref{eq:kQself})),
which gives, on an average, about 1 MeV for the excitation energy
of $\gamma$-vibrations in the above model space.
We believe that this choice is more suitable
to understand the systematic behaviors of the result of calculation
for nuclei in the rare-earth region.

    One of the most important output quantities is the rotational
energy parameters, i.e. the Harris parameters, in our formalism of
the $\omega$-expansion.  Up to the third order,
\beq
  I_x(\omega)=I+1/2=\cJ_0 \,\omega + \cJ_1 \,\omega^3,
\label{eq:Harris}
\eeq
where $I=0,2,4,\cdots$ ($\hbar$)
for the $K=0$ ground state bands.\cite{BF79} \ 
The results are summarized in Table~\ref{tab:SCCsum}
in comparison with experimental data,
where the experimental Harris parameters $\cJ_0$ and $\cJ_1$ 
are extracted from the $E_{2^+}$ and $E_{4^+}$ of the ground state band
as follows:
\beq
  \cJ_0 =\frac{1.5\,\omega_2^3-3.5\,\omega_1^3}
  {\omega_1\omega_2^3-\omega_1^3\omega_2}, \quad
  \cJ_1 =\frac{3.5\,\omega_1-1.5\,\omega_2}
  {\omega_1\omega_2^3-\omega_1^3\omega_2},
\label{eq:expJ01}
\eeq
with
\beq
  \omega_1 \equiv E_{2^+}/2, \quad \omega_2 \equiv (E_{4^+}-E_{2^+})/2.
\label{eq:expJ01a}
\eeq
If the resultant parameter becomes negative
or $\cJ_1$ gets greater than 1000 $\hbar^4/$MeV$^3$,
then only $\cJ_0=3/E_{2^+}$ parameters are shown in Table~\ref{tab:SCCsum}.
It is seen from the table that two Harris parameters are nicely
reproduced, especially their mass number dependence.
In contrast to the $\cJ_0$ parameter, for which only the residual quadrupole
pairing interaction affects, the $\cJ_1$ parameter are sensitive
to all components of the residual interaction.
In other words, $\cJ_1$ reflects the mode-mode couplings
of the collective rotation to other elementary excitation modes.
Therefore the SCC method with the present residual interaction
is considered to be a powerful means
to describe the ``non-adiabaticity'' of nuclear collective rotations.
Details of coupling mechanism has been investigated in Ref.~\citen{TMS91}
by decomposing the contributions from various RPA eigenmodes:
It has been found that the couplings to the pairing vibrations
and collective surface vibrations are especially important.
Although the main contributions come from the collective modes,
many RPA eigenmodes have to be included
to reach the correct results, see also Ref.~\citen{MMgamma} for this point.
The method of the response-function matrix described in \S\ref{sec:SCCsol}
is very useful to include all RPA eigenmodes.

%%%%%%%%%%%
%  Table  %
%%%%%%%%%%%

\begin{table}
\caption{
  Results of the $\omega$-expansion for some observables 
  in Er ($Z=68$) isotopes.
  $Q^{(+)}_{2K}$ ($K=0,2$) are expectation values of the mass quadrupole
  operators.  The zero-th order values of $\Delta$ are shown
  in Table~\protect\ref{tab:SCCsum}, and those of $Q^{(+)}_{22}$ are zero
  (axially symmetric at $\omega=0$).  Units of each quantity are
  shown in the second raw.
  }
\label{tab:SCCexpd}
\begin{center}
\begin{tabular}{cccccccccccc}
\hline\hline
 & $N$ & $(N)_1$ & $(Z)_1$
 & $(\Delta_\nu)_1$ & $(\Delta_\pi)_1$
 & $(Q^{(+)}_{20})_0$ & $(Q^{(+)}_{20})_1$ & $(Q^{(+)}_{22})_1$ \\
 &     & $\hbar^2$/MeV$^2$ & $\hbar^2$/MeV$^2$
 & $\hbar^2$/MeV & $\hbar^2$/MeV
 & b & b$\,\hbar^2$/MeV$^2$ & b$\,\hbar^2$/MeV$^2$ \\
\hline
Er & 88 &   14.5 & $-$4.2 & $-$0.45 & $-$1.81 & 2.84 & 6.14 & 4.58 \\
   & 90 &    9.9 & $-$5.4 & $-$0.72 & $-$1.80 & 3.74 & 4.67 & 4.78 \\
   & 92 &    7.5 & $-$6.3 & $-$1.58 & $-$1.83 & 4.71 & 4.52 & 5.72 \\
   & 94 &    3.6 & $-$4.3 & $-$2.12 & $-$1.50 & 5.13 & 2.45 & 4.43 \\ 
   & 96 &    1.8 & $-$3.1 & $-$2.34 & $-$1.33 & 5.50 & 1.31 & 3.70 \\
   & 98 &    1.0 & $-$3.0 & $-$2.83 & $-$1.32 & 5.71 & 1.10 & 2.74 \\
   &100 & $-$2.6 & $-$3.4 & $-$2.83 & $-$1.35 & 5.82 & 0.86 & 1.51 \\
   &102 & $-$3.5 & $-$3.6 & $-$2.34 & $-$1.38 & 5.87 & 0.79 & 1.08 \\
\hline
\end{tabular}
\end{center}
\end{table}

%%%%%%%%%%%
%  Table  %
%%%%%%%%%%%

\begin{table}
\caption{
  Similar to Table~\protect\ref{tab:SCCexpd} but
  the residual interactions are artificially switched off in the calculation.
  The results for the Harris parameters are also included.
  }
\label{tab:SCCnoint}
\begin{center}
\begin{tabular}{cccccccccccc}
\hline\hline
 & $N$ & $\cJ_0$ & $\cJ_1$ & $(N)_1$
 & $(\Delta_\nu)_1$
 & $(Q^{(+)}_{20})_1$ & $(Q^{(+)}_{22})_1$ \\
 &     & $\hbar^2$/MeV & $\hbar^4$/MeV$^3$ & $\hbar^2$/MeV$^2$
 & $\hbar^2$/MeV
 & b$\,\hbar^2$/MeV$^2$ & b$\,\hbar^2$/MeV$^2$ \\
\hline
Er & 88 &  7.5 & 11 &    1.00 & $-$0.18 & 0.32 & 0.13 \\
   & 90 & 12.1 & 18 &    1.05 & $-$0.25 & 0.38 & 0.15 \\
   & 92 & 18.1 & 28 &    0.97 & $-$0.39 & 0.42 & 0.18 \\
   & 94 & 21.4 & 34 &    0.60 & $-$0.48 & 0.33 & 0.18 \\ 
   & 96 & 24.6 & 34 &    0.67 & $-$0.54 & 0.27 & 0.18 \\
   & 98 & 27.6 & 51 &    0.83 & $-$0.67 & 0.29 & 0.17 \\
   &100 & 28.3 & 55 & $-$1.01 & $-$0.67 & 0.15 & 0.15 \\
   &102 & 27.5 & 36 & $-$0.85 & $-$0.57 & 0.16 & 0.15 \\
\hline
\end{tabular}
\end{center}
\end{table}

    Expectation values of other observable quantities are also
expanded in power series of $\omega$, and their coefficients
give us important information about the response of nucleus
against the collective rotation.  In Table~\ref{tab:SCCexpd}
we show examples for the nucleon number, monopole-pairing gaps, and
mass quadrupole moments:
\ben
  \braketc{\phi_\intr(\omega)}{N_\tau}{\phi_\intr(\omega)}
  &=& (N_\tau)_0 + (N_\tau)_1\,\omega^2,
\label{eq:Nexpd}  \\
  G_0^\tau\braketc{\phi_\intr(\omega)}{P^\tau_{00}}{\phi_\intr(\omega)}
  &=& (\Delta_\tau)_0 + (\Delta_\tau)_1\,\omega^2,
\label{eq:D00expd}  \\
  \braketc{\phi_\intr(\omega)}{Q^{(+)}_{2K}}{\phi_\intr(\omega)}
  &=& (Q^{(+)}_{2K})_0 + (Q^{(+)}_{2K})_1\,\omega^2 \;\;\; (K=0,2).
\label{eq:Qexpd}
\een
They are time-reversal even quantities so that
the series contains up to the second order within the third order calculations.
It should be noticed that these $\omega$-expanded quantities
are associated with the properties of the diabatic ground state band,
which becomes non-yrast after the $g$-$s$ band-crossing.
As remarked in the end of \S\ref{sec:SCCsol},
$(N_\tau)_1 \ne 0$ means that the nucleon number is not conserved
along the rotational band.  However, its breakdown is rather small;
even in the worst case of $^{156}$Er in Table~\ref{tab:SCCexpd}
the deviation is about 1.3 at $\omega=0.3$ MeV, and it is less than
0.1 at $\omega=0.1$ MeV in $^{166}$Er.
It is well known that the pairing gap decreases as a function of $\omega$
due to the Coriolis anti-pairing effect.
It is sometimes phenomenologically parametrized as\cite{WSNJ90}
\beq
  \Delta(\omega)=\left\{
  \bea{ll}
      \Delta_0 \Bigl( 1 - {\displaystyle \frac{1}{2}
        \bigl(\frac{\omega}{\omega_{\rm c}}\bigr)^2} \Bigr)
    & \omega \le \omega_{\rm c}, \\
      {\displaystyle \frac{1}{2}\Delta_0
        \bigl(\frac{\omega_{\rm c}}{\omega}\bigr)^2}
    & \omega > \omega_{\rm c}. \\
  \eea \right.
\label{eq:phenomDel}
\eeq
Thus, our $\omega$-expansion method precisely gives
the phenomenological parameter
$\omega_{\rm c} = \sqrt{-\Delta_0/2\Delta_1}$ ($\Delta_1 < 0$)
in Eq.~(\ref{eq:phenomDel}) from microscopic calculations.
As shown in Table~\ref{tab:SCCexpd},
$(\Delta_\nu)_1$ varies considerably along the isotopic chain. 
The $(Q_{2K})_1$ are related to the shape change at high-spin states,
and tell us how soft the nucleus is against rapid rotation. 
Since nuclei studied in the present work are axially symmetric
in their ground states, $(Q_{20})_1$ and $(Q_{22})_1$ serve as measures
of softness in the $\beta$- and $\gamma$-directions, respectively.
As seen in Table~\ref{tab:SCCexpd} the isotopes
get harder in both directions as the neutron number increases;
especially, the $N=88$ and $N=90$ isotopes are known to undergo
a shape change from the prolate collective to the oblate non-collective
rotation scheme at very high-spin states (``band termination''\cite{AFLR99}),
while heavier isotopes ($N \ge 96$) are known to be well deformed
keeping prolate shape until the highest observed spins.
These features have been well known from the calculations of
the potential energy surface in the $(\epsilon_2,\gamma)$-plane,
and our results seem to agree with them qualitatively.
In order to see the effect of the residual interactions,
the result obtained by neglecting them, i.e. that of a simple higher order
Coriolis coupling calculations, is shown in Table~\ref{tab:SCCnoint}.
Comparing it with Tables~\ref{tab:SCCsum} and~\ref{tab:SCCexpd},
it is clear that the residual interactions play an important role
in the $\omega$-dependence of observables.
For example, $\cJ_1$ Harris parameter becomes quite small
by a factor of about 1/2$-$1/3 when the residual interactions are switched off.
The effect on the second order coefficients of the quadrupole moment
is more dramatic and leads to about an order of magnitude reduction
in soft nuclei.

% **********
% * Figure
% **********
\begin{figure}
\epsfysize=6.1cm
\centerline{\epsfbox{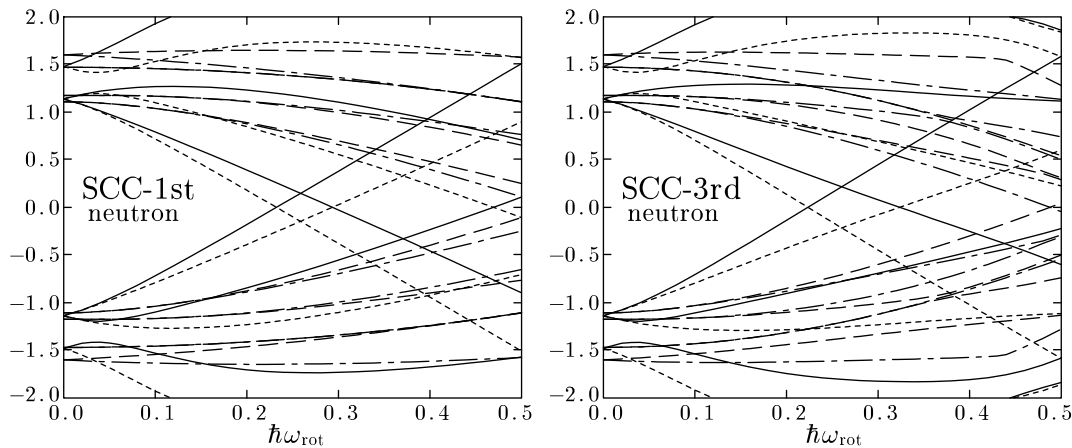}}
\caption{
 Neutron quasiparticle Routhians plotted 
 as functions of $\hbar \omega_\rot$ (MeV) suitable for $^{162}$Er.
 They are obtained by diagonalizing
 the SCC quasiparticle Hamiltonian~(\protect\ref{eq:DiabQPh})
 up to the first order (left) and third order (right)
 of the $\omega$-expansion.  As in the case of the usual adiabatic
 quasiparticle energy diagram, the negative energy solutions,
 $-E'_\mu=E'_{\bar \mu}$ and $ -E'_{\bar \nu}=E'_\nu$, are also drawn.
 The solid,  dotted, dashed, and dash-dotted curves denote Routhians
 with $(\pi,r)=(+,+i)$, $(+,-i)$, $(-,+i)$, and $(-,-i)$, respectively.
  } 
\label{fig:routhnscc}
\end{figure}

% **********
% * Figure
% **********
\begin{figure}
\epsfysize=6.1cm
\centerline{\epsfbox{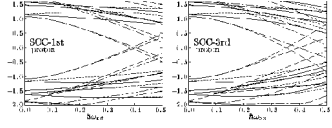}}
\caption{
 Same as Fig.~\protect\ref{fig:routhnscc} but for proton quasiparticles.
  } 
\label{fig:routhpscc}
\end{figure}

% **********
% * Figure
% **********
\begin{figure}
\epsfysize=6.1cm
\centerline{\epsfbox{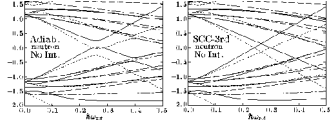}}
\caption{
 Same as Fig.~\protect\ref{fig:routhnscc} but obtained
 by the adiabatic cranking (left) and the third order SCC (right)
 with neglecting the residual interactions.
  } 
\label{fig:routhnni}
\end{figure}

% **********
% * Figure
% **********
\begin{figure}
\epsfysize=8.7cm
\centerline{\epsfbox{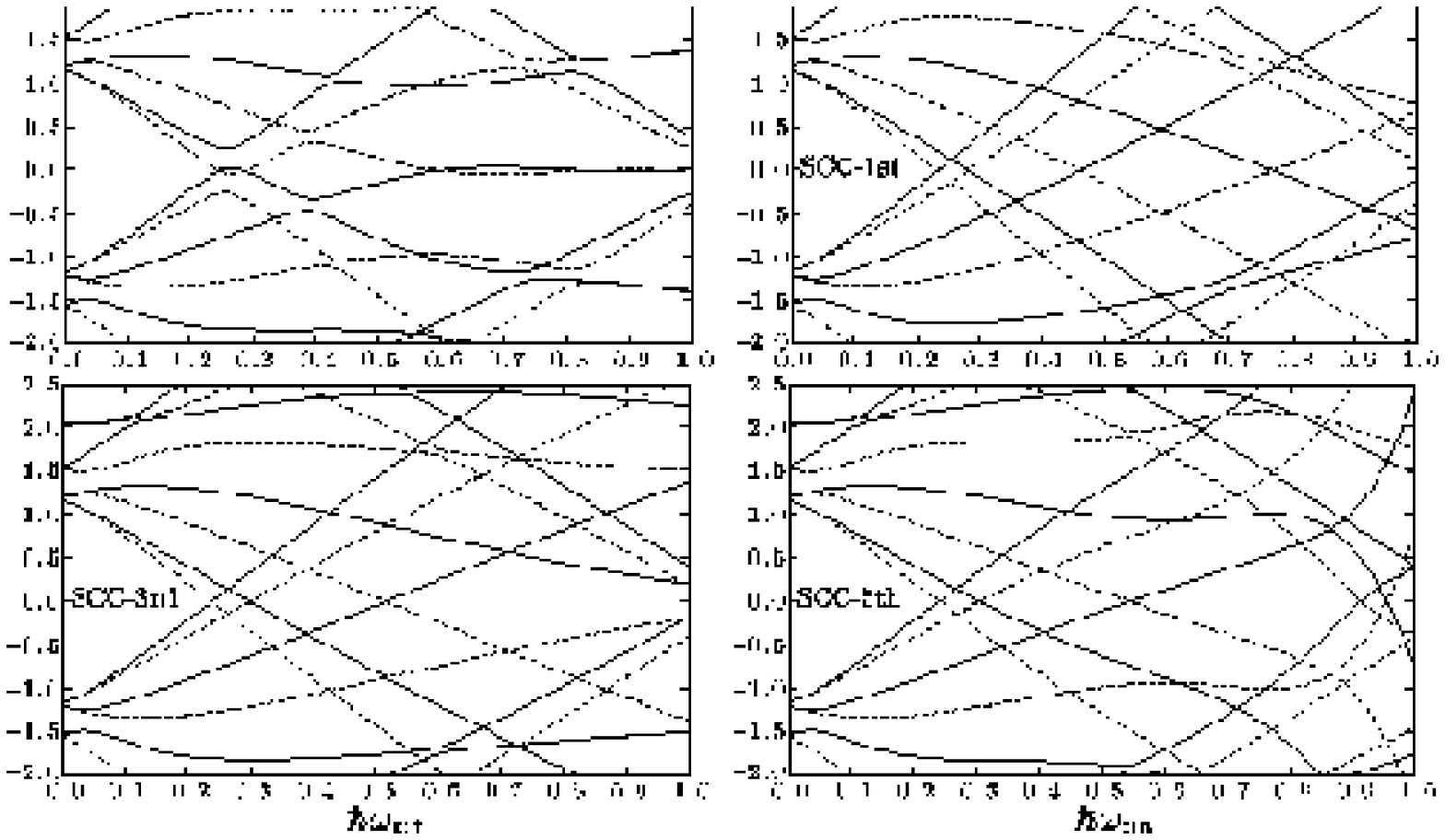}}
\caption{
 Neutron quasiparticle Routhians for the $N_\osc=6$ ($i_{13/2}$) orbits
 suitable for $^{162}$Er.
 The left upper, right upper, left lower, and right lower
 panels denote the results of the adiabatic cranking,
 the SCC up to the first order, 3rd order, and 5th order, respectively.
 The solid and dotted curves denote Routhians with
 $r=+i$ $(\alpha=-1/2)$ and $r=-i$ $(\alpha=+1/2)$, respectively.
 The effect of residual interaction is completely neglected
 in this calculation.
  } 
\label{fig:routhnall}
\end{figure}

    Now let us study the quasiparicle Routhians obtained by means of 
the SCC method.
It is mentioned in \S\ref{sec:SCCqpe} that the two step diagonalization
with the truncation of the $\omega$-expansion up to $\nmax$,
c.f. Eq.~(\ref{eq:DiabQPh}), leads to
diabatic quasiparticle states in the rotating frame,
in which the negative and positive eigenstates do not interact with each other.
We show in Figs.~\ref{fig:routhnscc} and~\ref{fig:routhpscc}
calculated quasiparticle Routhians for neutrons and protons,
respectively.
It is  confirmed that the diabatic quasiparticle states are obtained.
As discussed in \S\ref{sec:SCCqpe}, the diagonalization
of the quasiparticle Hamiltonian in the SCC method is completely
equivalent to that of the selfconsistent cranking model,
which is known to lead to the adiabatic levels,
if the first step unitary transformation $e^{iG(\omega)}$
is treated non-perturbatively in full order.
Then what is the mechanism that realizes the diabatic levels?
We believe that the cutoff of the $\omega$-expansion extracts
the smoothly varying part of the quasiparticle Hamiltonian; namely,
ignoring its higher order terms eliminates the cause of
abrupt changes of the microscopic internal structure by
quasiparticle alignments.
An analogous mechanism has been known for many years
in the Strutinsky smoothing procedure:\cite{Brack72} \ 
The $\delta$-function in the microscopic level density
is replaced by the Gaussian smearing function times
the sum of the Hermite polynomials (complete set),
and the lower order cutoff of the sum
(usually 6th order is taken) gives the smoothed level density.
It should be noted, however, that the plateau condition guarantees
that the order of cutoff does not affect the physical results
in the case of the Strutinsky method.
We have not yet succeeded in obtaining such a condition
in the present case of the cutoff of the $\omega$-expansion
in the SCC method for the collective rotation.
Therefore we have to decide the $\nmax$ value by comparison
of the calculated results with experimental data.
We mainly take $\nmax=3$ in the following;
determination of the optimal choice of $\nmax$
remains as a future problem.

    In Figs.~\ref{fig:routhnscc} and~\ref{fig:routhpscc}
the results obtained by truncating
up to the first order ($\nmax=1$) and the third order ($\nmax=3$)
are compared.
It is clear that the higher order terms considerably modify
the quasiparticle energy diagrams.
Especially, the alignments of the lowest pair of quasiparticles
are reduced for neutrons (low $K$ states of the $i_{13/2}$-orbitals),
while they are increased for protons
(medium $K$ states of the $h_{11/2}$-orbitals).
Thus, the higher order effects depend strongly on the nature of orbitals.
It should be stressed that the effects of the residual interaction,
i.e. changes of the mean-field against the collective rotation,
are contained in the quasiparticle diagrams presented in these figures.
In this sense, they are different from the spectra of the cranked
shell model,\cite{BF79} \ where the mean-field is fixed at $\omega=0$.
In Fig.~\ref{fig:routhnni} are displayed the usual
adiabatic quasineutron Routhians and the third order SCC Routhians,
in both of which the residual interactions are neglected completely.
Again, by comparing Fig.~\ref{fig:routhnni}
with Fig.~\ref{fig:routhnscc},
it is seen that the effect of residual interactions
considerably changes the quasiparticle states.
In relation to the choice of $\nmax$, we compare in Fig.~\ref{fig:routhnall}
the Routhians obtained by changing the cutoff order $\nmax=1,3,5$.
In this figure, the usual non-selfconsistent adiabatic Routhians are
also displayed, and for comparison's sake, the residual interactions
are completely neglected in all cases.
Moreover, the rotational frequency is extended to unrealistically
large values in order to see the asymptotic behaviors of the Routhian.
Comparing the adiabatic Routhians with those of the SCC method,
positive and negative energy solutions cross irrespective of
the strength of level-repulsion.
Although the adiabatic levels change their characters
abruptly at the crossing,
if their average behaviors are compared to the calculated ones,
the third order results ($\nmax=3$) agree best with the adiabatic levels.
The first order results, for example, give the alignments
(the slopes of Routhians) too large.
On the other hand, the divergent behaviors are clearly seen
at about $\omega \ge 0.8$ MeV in the fifth order results.
The inclusion of the effect of the residual interactions
makes this convergence radius in $\omega$ even smaller.

% **********
% * Figure
% **********
\begin{figure}
\epsfysize=8.1cm
\centerline{\epsfbox{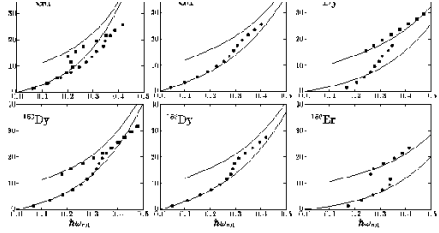}}
\caption{
 Comparison of the third order SCC method calculations
 for the diabatic $g$- and $s$-bands with experimental data.
 The angular momenta $\braketa{J_x}=I+1/2$ ($\hbar$) are displayed versus
 the rotational frequency $\hbar\omega_\rot$ (MeV) for nuclei
 in the rare-earth region, Gd ($Z=64$) to W ($Z=74$) isotopes.
 Filled circles denote experimental data smoothly extended from
 the ground state.  Data for excited bands are also included
 as filled squares when available, which are, in most cases, identified
 as $s$-bands.
  }
\label{fig:SCCgsall}
\end{figure}

% **********
% * Figure
% **********
\addtocounter{figure}{-1}
\begin{figure}
\epsfysize=8.1cm
\centerline{\epsfbox{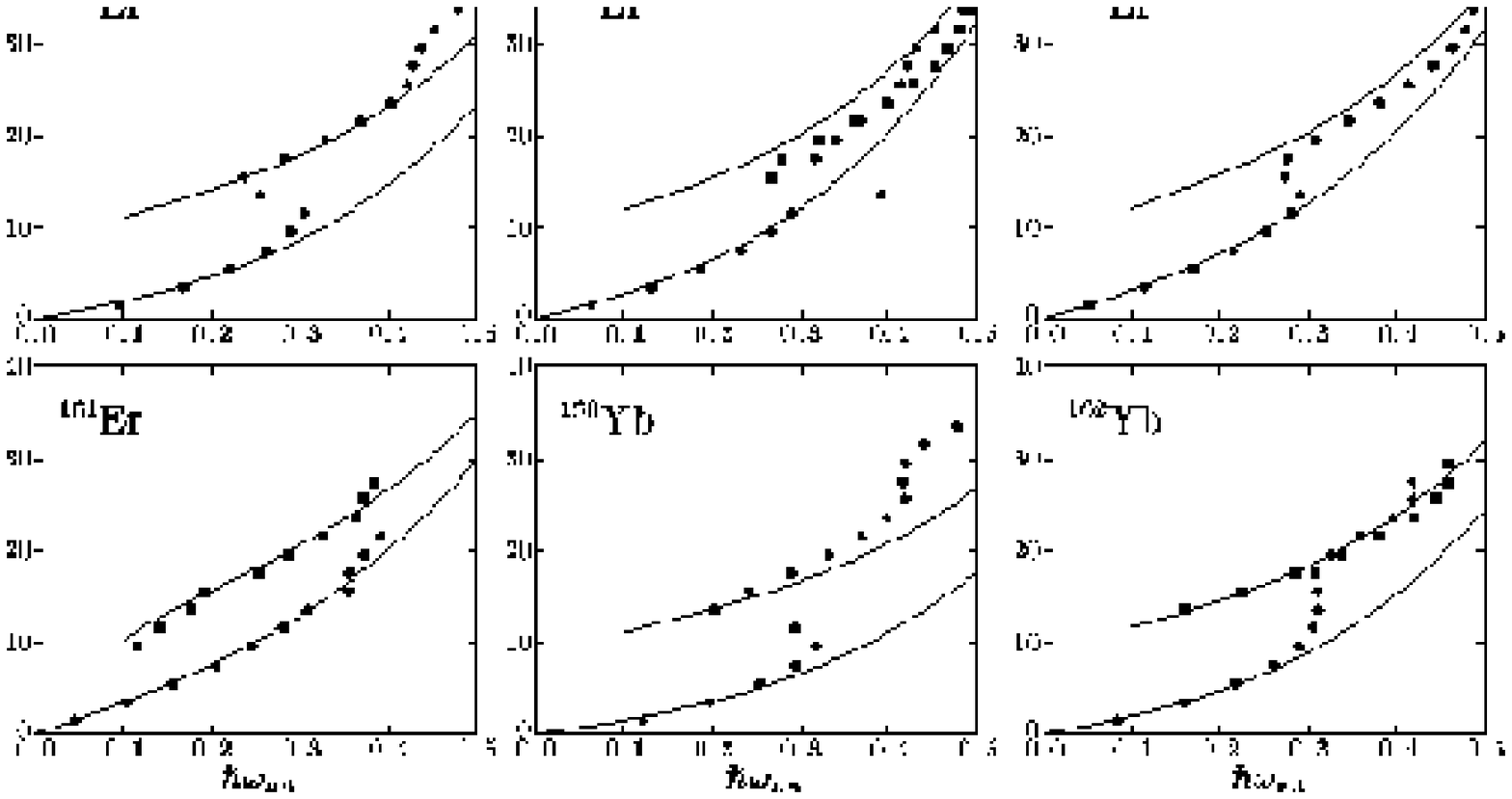}}
\caption{
 {\it continued.}
  }
\end{figure}

% **********
% * Figure
% **********
\addtocounter{figure}{-1}
\begin{figure}
\epsfysize=8.1cm
\centerline{\epsfbox{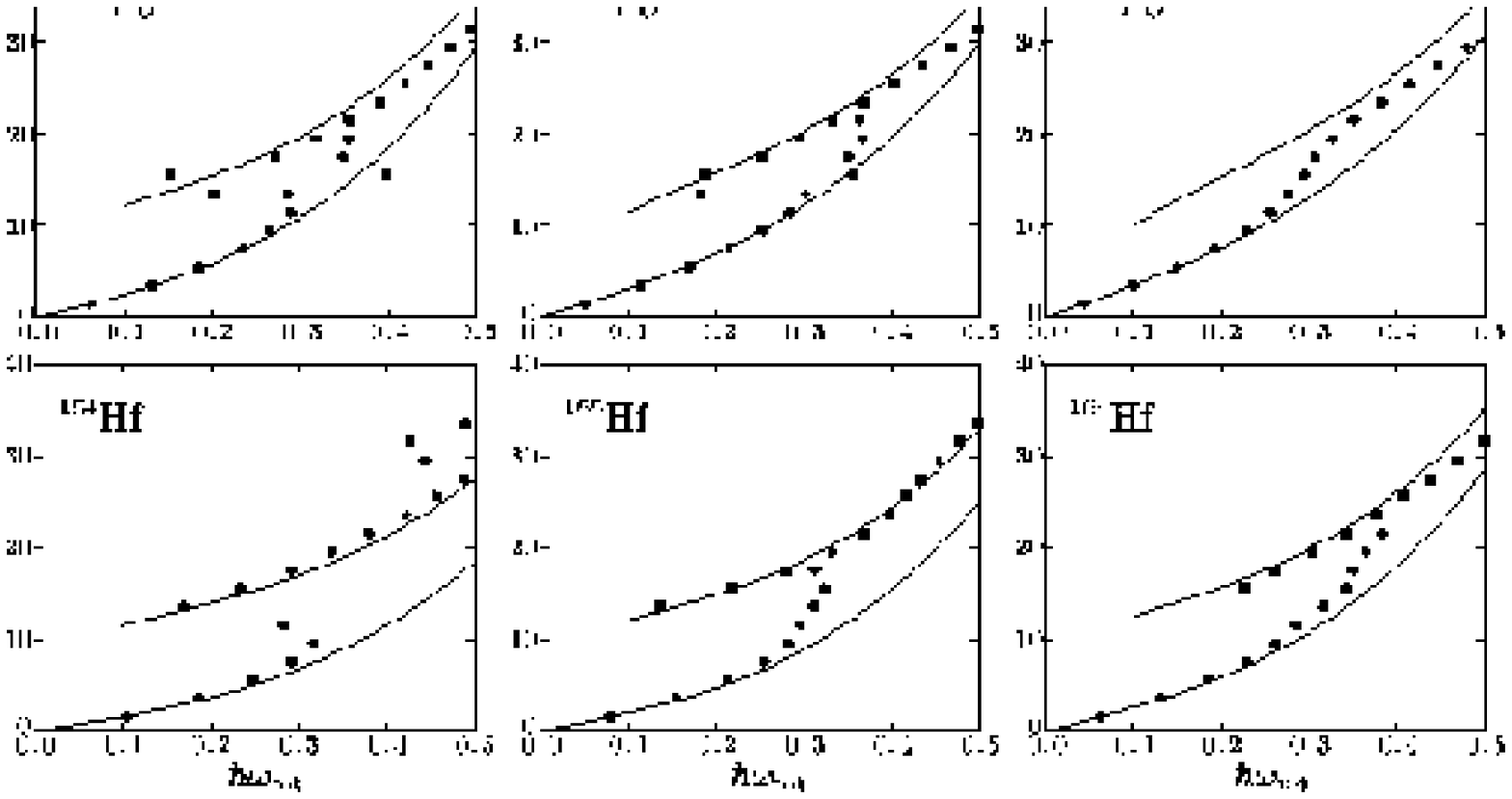}}
\caption{
 {\it continued.}
  }
\end{figure}

% **********
% * Figure
% **********
\addtocounter{figure}{-1}
\begin{figure}
\epsfysize=8.1cm
\centerline{\epsfbox{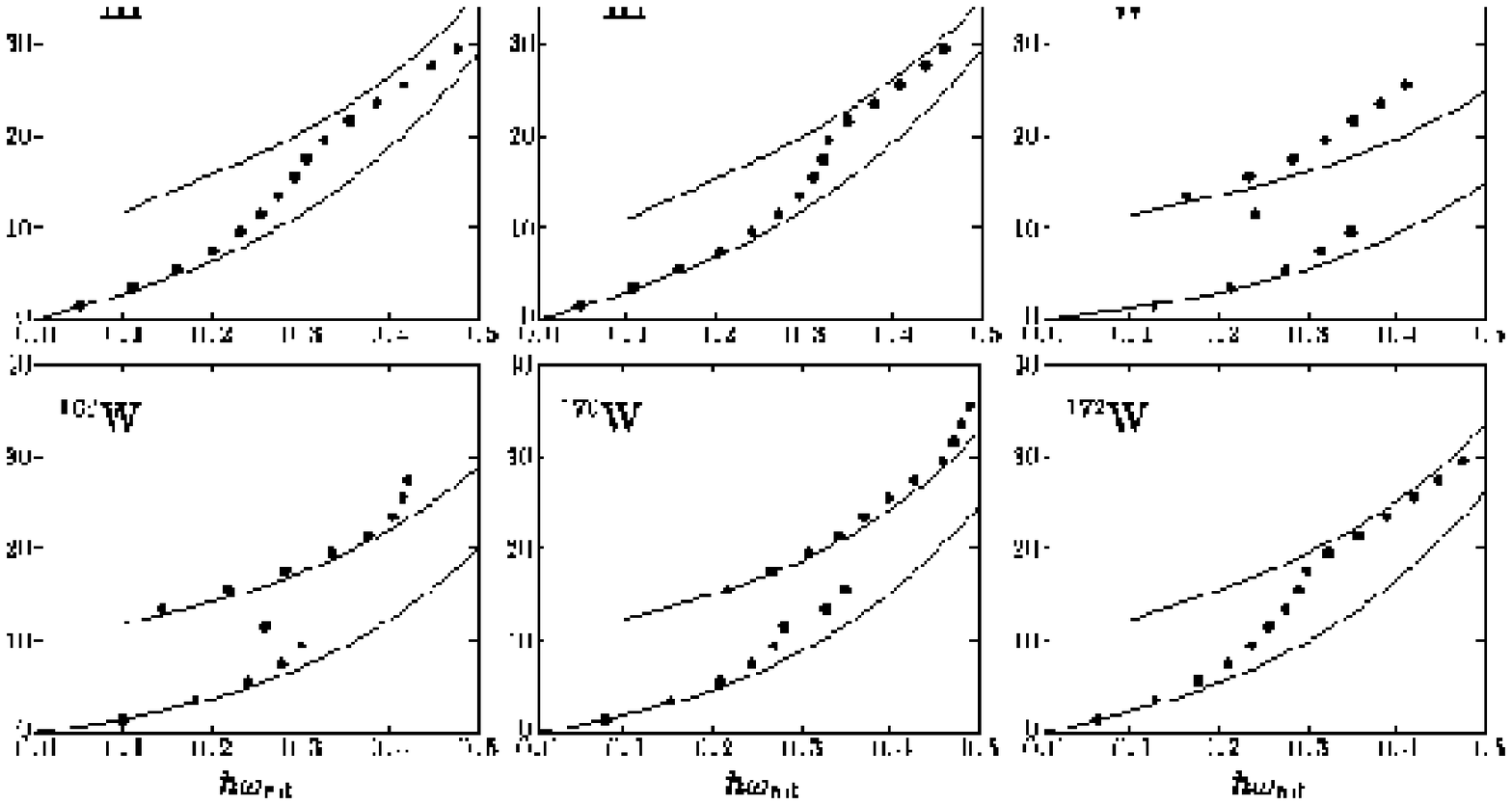}}
\caption{
 {\it continued.}
  }
\end{figure}

    Finally, we would like to discuss the results of application
of the present formalism to the $g$- and $s$-bands, which are observed
systematically and compose the yrast lines of even-even nuclei.  
Although we can compare the Routhians~(\ref{eq:colRouth}),
or equivalently the rotational energy~(\ref{eq:colH}),
it is known that the relation
$I_x$ versus $\omega$ gives a more stringent test.
Therefore we compare the calculated $I_x-\omega$ relation
with the experimental one in Fig.~\ref{fig:SCCgsall}
for even-even nuclei in the rare-earth region,
in which the band crossings are identified along the yrast sequences.
In this calculation the $I_x(\omega)$ of the $g$-band is
given by Eq.~(\ref{eq:Harris}) with calculated values
of the Harris parameters (see Table~\ref{tab:SCCsum}).
As for the $I_x(\omega)$ of the $s$-band, we calculate it
on the simplest assumption of the independent quasiparticle motions
in the rotating frame, which is the same as that of the cranked shell model:
\beq
  \ket{\phi_s(\omega)}=\alpha^\dagger_1(\omega)
           \alpha^\dagger_{\bar 1}(\omega) \ket{\phi_\intr(\omega)},
  \quad \ket{\phi_g(\omega)}=\ket{\phi_\intr(\omega)},
\label{eq:gsWaveFun}
\eeq
where $\alpha^\dagger_1(\omega)$ and $\alpha^\dagger_{\bar 1}(\omega)$
are the lowest $r=+i$ and $r=-i$ quasineutron creation operators
in the rotating frame.  Then, the $I_x(\omega)$ of the $s$-band
is the sum of $I_x(\omega)$ of the $g$-band and the aligned
angular momenta of two quasineutrons, which are calculated according to
Eqs.~(\ref{eq:OrotRep})$-$(\ref{eq:OrotA-}),
\beq
   \bigl(I_x(\omega)\bigr)_{s\hbox{-}{\rm band}} = I_x(\omega)
   + i_1(\omega)+i_{\bar 1}(\omega), \quad
   \bigl(I_x(\omega)\bigr)_{g\hbox{-}{\rm band}} = I_x(\omega),
\label{eq:Iofgs}
\eeq
or by using the canonical relation between
the Routhian and the aligned angular momentum,
the alignments $i_\mu$ and $i_{\bar \mu}$ can be calculated as usual:
\beq
   i_\mu(\omega)=- \frac{\partial E'_\mu(\omega)}{\partial \omega},
   \quad
   i_{\bar \mu}(\omega)=
   - \frac{\partial E'_{\bar \mu}(\omega)}{\partial \omega}.
\eeq
Since our quasiparticle Routhians behave diabatically as functions
of the rotational frequency, the resultant $g$- and $s$-bands
are also non-interacting bands; we have to mix them at the same angular
momentum to obtain the interacting bands
corresponding to the observed bands.
Such a band mixing calculation is straightforward in our formalism
if the interband $g$-$s$ interaction is provided.
However, it is a very difficult task
as long as the usual adiabatic cranking model is used.
In the present stage we are not able to estimate the $g$-$s$ interband
interaction theoretically.
Therefore, we do not attempt to perform such band-mixing calculations
in the present paper (but see \S\ref{sec:gsInt}).

    Looking into the results displayed in Fig.~\ref{fig:SCCgsall}, one see
that our diabatic formalism of collective rotation based on
the SCC method is quite successful.  The overall agreements are
surprisingly good, considering the fact that
we have only used a global parametrization of the strengths
of the residual interaction:
\ben
  &&G_0^\nu=20/A, \quad G_0^\pi=24/A \;\;{\rm (MeV)},
\label{eq:strengthG0}  \\
  &&g_2^\nu=g_2^\pi=24,
\label{eq:strengthG2}  \\
  &&\kappa_{2K}=1.45\,\kappa_2^\self,
\label{eq:strengthK2}
\een
for the model space of three $N_\osc$-shells
($4-6$ for neutrons and $3-5$ for protons).
The agreements of the calculated $g$-bands come from the fact that
the Harris parameters (Table~\ref{tab:SCCsum}) are nicely
reproduced in the calculation.   Further agreements of
the $s$-bands are not trivial, and tell us that we have
obtained reliable diabatic quasiparticle spectra
(Figs.~\ref{fig:routhnscc} and~\ref{fig:routhpscc}).
It is known that, if the $I_x-\omega$ relations of $s$-bands are
parametrized in the form, $I_x = i+\cJ_0\,\omega+\cJ_1\,\omega^3$,
the $\cJ_1$ Harris parameters of $s$-bands are systematically
smaller than those of $g$-bands.
This feature is quite well reproduced in the calculations,
as is clearly seen in Fig.~\ref{fig:SCCgsall},
and the reason is that the value of the aligned angular momentum
of two quasineutrons decreases as a function of $\omega$.
The suitable decrease is obtainable only if the residual interactions
are included and the diabatic quasiparticle Routhians are evaluated up to
the third order.

% ***********************************
% * SECTION 4
% ***********************************
\section{ Diabatic quasiparticle basis
and the interband interaction between the ${\mib g}$- and ${\mib s}$-bands}
\label{sec:SCCDQPbasis}

   The formulation of the previous section gives a consistent
perturbative solution, with respect to the rotational frequency, 
of the basic equations of the SCC method for collective rotation.
However, it has a problem
as a method to construct the diabatic quasiparticle basis:
The wave functions of the diabatic levels are orthonormal
only within the order of cutoff ($\nmax$) of the $\omega$-expansion.
In the previous section only the independent quasiparticle states,
i.e. one-quasiparticle states or the $g$- and $s$-bands,
are considered and this problem does not show up.
The quasiparticle states have another important role that
they are used as a basis of complete set for a more sophisticated
many body technique beyond the mean-field approximation;
for example, the study of collective vibrations at high spin 
in terms of the RPA method 
in the rotating frame.\cite{SM83,MSM88,Mar77,JM79,EMR80,Zel80} \ 
In such an application it is crucial that
the diabatic quasiparticle basis satisfies the orthonormal property.
We present in this section a possible method to construct the diabatic basis
satisfying the orthonormality condition. 

   Another remaining problem which is not touched in the previous
section is how to theoretically evaluate the interband interaction
between the ground state band and the two quasineutron aligned band.
Since we do not have satisfactory answer yet to this problem,
we only present a scope for possible solutions at the end of this section.

\subsection{ Construction of diabatic quasiparticle basis in the SCC method}
\label{sec:constDQPbasis}

   Although the basic idea is general, we restrict ourselves to the case
of collective rotation and use the good signature representation
with real phase convention, introduced in \S\ref{sec:SCCsol},
for the matrix elements of the Hamiltonian $H$ and 
of the angular momentum $J_x$.
First let us recall that the diabatic quasiparticle basis
in the rotating frame is obtained by
the two step unitary transformation~(\ref{eq:rotQPop}).
The first transformation by $e^{iG(\omega)}$ can be represented
as follows,\cite{MMSK80} \ 
\ben
  e^{iG(\omega)}  a^\dagger_i e^{-iG(\omega)}
  &=& \sum_{j>0}
     \bigl[\cos{\sqrt{g g^\T}}\bigr]_{ij} a^\dagger_j
   - \sum_{j>0}
   \Bigl[g\,{\displaystyle \frac{\sin{\sqrt{g^\T g}}}{\sqrt{g^\T g}} }
      \Bigr]_{ij} a_{\bar j},
\label{eq:QPtranGa} \\
  e^{iG(\omega)}  a^\dagger_{\bar i} e^{-iG(\omega)}
  &=& \sum_{j>0}
     \bigl[\cos{\sqrt{g^\T g}}\bigr]_{ij} a^\dagger_{\bar j}
   + \sum_{j>0}
   \Bigl[g^\T\,{\displaystyle \frac{\sin{\sqrt{g g^\T}}}{\sqrt{g g^\T}} }
      \Bigr]_{ij} a_j,
\label{eq:QPtranGb}
\een
with real matrix elements 
$g_{ij}(\omega)=\sum_{n \ge 1} \omega^n g^{(n)}(ij)$,
see Eq.~(\ref{eq:OpGnform}), where $g^\T$ denotes the transpose of $g$.
Thus, by using an obvious matrix notation, the transformation
to the rotating quasiparticle operator
from the $\omega=0$ quasiparticle operator is given as
\ben
  \left( \bea{l} \alpha \\ {\bar \alpha}^\dagger \eea \right)
  &=& \left( \bea{cc} f^\T & 0 \\ 0 & {\bar f}^{\,\T} \eea \right)
   \left( \bea{cc}
    \cos{\sqrt{g g^\T}} &
   -g\,{\displaystyle \frac{\sin{\sqrt{g^\T g}}}{\sqrt{g^\T g}} } \\
   g^\T\,{\displaystyle \frac{\sin{\sqrt{g g^\T}}}{\sqrt{g g^\T}} } &
     \cos{\sqrt{g^\T g}} \eea \right)
  \left( \bea{l}  a \\ {\bar a}^\dagger \eea \right)
\label{eq:QPtraT} \\
  &\equiv& \cF^\T(\omega) \,\cG^\T(\omega)
  \left( \bea{l}  a \\ {\bar a}^\dagger \eea \right),
\label{eq:QPtraTd}
\een
where the real matrix elements 
$f_{i\mu}(\omega)$ and ${\bar f}_{i\mu}(\omega)$
are the amplitudes that diagonalize
the quasiparticle Hamiltonian in the rotating frame,
see Eq.~(\ref{eq:eigQPE}), for signature $r=+i$ and $-i$, respectively.
The cutoff of the $\omega$-expansion means that
the generator $iG(\omega)$, i.e. the matrix $g$,
is solved up to the $n=\nmax$ order,
\beq
  g(\omega) = [g(\omega)]^{(n \le \nmax)}
  = \sum_{n=1}^\nmax \omega^n g^{(n)},
\label{eq:gcutoff}
\eeq
and at the same time the transformation matrix $\cG(\omega)$
itself is treated perturbatively
\beq
  \cG(\omega)
  = \left( \bea{cc}
  1 - \omega^2 g^{(1)}g^{(1)\T} + \cdots & \omega g^{(1)} + \cdots \\
  -\omega g^{(1)\T} + \cdots & 1 - \omega^2 g^{(1)\T}g^{(1)} + \cdots
   \eea \right),
\label{eq:Gtraexpd}
\eeq
while the other one, $\cF(\omega)$, is treated non-perturbatively by
the diagonalization procedure.
The origin of difficulty arising when the diabatic basis is utilized 
as a complete set lies in this treatment of $\cG(\omega)$,
because the orthogonality of the matrix $\cG(\omega)$
is broken in higher-orders.

    Now the solution to this problem is apparent:
The generator matrix $g(\omega)$ is solved perturbatively
like in Eq.(\ref{eq:gcutoff}),
but the transformation matrix $\cG(\omega)$ has to be treated
non-perturbatively as in Eq.~(\ref{eq:QPtraT}).
In order to realize this treatment we introduce new
orthogonal matrices, $D$ and ${\bar D}$,
which diagonalize $g g^\T$ and $g^\T g$ within the signature $r=+i$
and $-i$ states, respectively,
\beq
  \sum_{j>0} (g g^\T )_{ij} D_{jk} = D_{ik} \theta^2_k, \quad
  \sum_{j>0} (g^\T g)_{ij} {\bar D}_{jk} = {\bar D}_{ik} \theta^2_k,
\label{eq:gdiagD}
\eeq
where we have used the fact that the matrices $g g^\T$ and $g^\T g$
have common eigenvalues, which are non-negative, and then we have
\ben
  \cG(\omega) = \left( \bea{cc}
  D (\cos{\theta}) D^\T & g {\bar D} (\sin{\theta}/\theta) {\bar D}^\T \\
  -g^\T D (\sin{\theta}/\theta) D^\T & {\bar D} (\cos{\theta}) {\bar D}^\T
   \eea \right).
\label{eq:QPtraTdiagD}
\een
Here ($\cos{\theta}$) and ($\sin{\theta}/\theta$) denote diagonal matrices,
whose matrix elements are $\delta_{ij}\cos{\theta_i}$
and $\delta_{ij}\sin{\theta_i}/\theta_i$, respectively.
The physical meaning is that the orthogonal matrices $D$ and ${\bar D}$
are transformation matrices from the quasiparticle operators
$(a^\dagger_i, a_i)$ and $(a^\dagger_{\bar i}, a_{\bar i})$
at $\omega=0$ to their canonical bases, which diagonalize
the density matrices $\rho$ and ${\bar \rho}$
with respect to the rotational HB state $\ket{\phi_\intr(\omega)}$,
respectively;
\ben
   \rho_{ij} &\equiv&
   \braketc{\phi_\intr(\omega)}{a^\dagger_i a_j}{\phi_\intr(\omega)}
  = \bigl[\cos{\sqrt{g g^\T}} \bigr]_{ij},
          \nonumber  \\
   {\bar \rho}_{ij} &\equiv&
   \braketc{\phi_\intr(\omega)}
    {a^\dagger_{\bar i} a_{\bar j}}{\phi_\intr(\omega)}
  = \bigl[\cos{\sqrt{g^\T g}} \bigr]_{ij}.
\label{eq:denmat}
\een
Thus the method to construct the rotating quasiparticle basis is summarized
as follows. First, solve the basic equation of the SCC method and
obtain the generator matrix $g(\omega)$ up to the $\nmax$ order
as in Eq.~(\ref{eq:gcutoff}).
At the same time, diagonalize the quasiparticle Hamiltonian
and obtain the eigenstates as in Eq.~(\ref{eq:eigQPE}) for
both signatures $r=\pm i$.
Secondly, diagonalize the density matrices~(\ref{eq:denmat}),
or equivalently Eq.~(\ref{eq:gdiagD}), and obtain
the orthogonal matrices $D$ and ${\bar D}$ of the canonical bases.
Finally, by using these matrices $D$ and ${\bar D}$ calculate
the transformation matrix $\cG(\omega)$ as in Eq.~(\ref{eq:QPtraTdiagD}),
and then the basis transformation is determined by Eq.~(\ref{eq:QPtraTd}).

   It is instructive to consider a concrete case of the cranked shell model;
i.e. the effect of residual interactions or the selfconsistency of mean-field
is neglected at $\omega>0$.  The quasiparticle basis is obtained by
diagonalizing the generalized Hamiltonian matrix:
\beq
  \left( \bea{cc} h_\Nils - \omega j_x & -\Delta \\
  -\Delta & -(h_\Nils + \omega j_x) \eea \right)
  \left( \bea{cc} U & {\bar V} \\ V & {\bar U} \eea \right)
 = \left( \bea{cc} U & {\bar V} \\ V & {\bar U} \eea \right)
  \left( \bea{cc} E' & 0 \\ 0 & -{\bar E}' \eea \right),
\label{eq:diagCSM}
\eeq
where $h_\Nils$ and $j_x$ denote matrices with respect to the Nilsson
(or the harmonic oscillator) basis at $\omega=0$,
and $(U,V)$ and $({\bar U},{\bar V})$ are coefficients of
the generalized Bogoliubov transformations from the Nilsson nucleon
operators $(c^\dagger_i, c_i)$ and $(c^\dagger_{\bar i}, c_{\bar i})$
(in the good signature representation),
\beq
  \alpha^\dagger_\mu =
  \sum_{i>0} (U_{i\mu} c^\dagger_i +V_{i\mu} c_{\bar i}), \quad
  \alpha^\dagger_{\bar \mu} =
  \sum_{i>0} ({\bar U}_{i\mu} c^\dagger_{\bar i} +{\bar V}_{i\mu} c_i),
\label{eq:QPtraCSM}
\eeq
or in the matrix notation
\beq
  \left( \bea{l} c \\ {\bar c}^\dagger \eea \right)
  = \cU \left( \bea{l} \alpha \\ {\bar \alpha}^\dagger \eea \right), \quad
  \cU \equiv \left( \bea{cc} U & {\bar V} \\ V & {\bar U} \eea \right).
\label{eq:QPtraCSMU}
\eeq
In contrast, the transformation $\cU$ is decomposed into three steps
in our construction method of the diabatic quasiparticle basis; 
(i) the Bogoliubov transformation $\cU_0$
between the nucleon $(c, {\bar c}^\dagger)$ and
the quasiparticle $(a, {\bar a}^\dagger)$ at $\omega=0$,
\beq
  \left( \bea{l} c \\ {\bar c}^\dagger \eea \right)
  = \cU_0 \left( \bea{l} a \\ {\bar a}^\dagger \eea \right), \quad
  \cU_0 \equiv \left( \bea{cc} u & v \\ -v^\T & u \eea \right),
\label{eq:QPtraU0}
\eeq
where $u$ and $v$ are the matrices of transformation at $\omega=0$,
(they are diagonal, e.g. $u_{ij} = u_i \delta_{ij}$, if only
the monopole-pairing interaction is included),
(ii) the transformation matrix $\cG(\omega)$ in Eq.~(\ref{eq:QPtraTd}),
generated by $e^{iG(\omega)}$, and (iii) the diagonalization step
of the rotating quasiparticle Hamiltonian $\cF(\omega)$
in Eq.~(\ref{eq:QPtraTd}), see also Eq.~(\ref{eq:eigQPE}), namely
\beq
  \cU(\omega)^{\rm SCC} = \cU_0 \,\cG(\omega) \cF(\omega).
\label{eq:QPtraSCCU}
\eeq
Here both $\cG(\omega)$ and $\cF(\omega)$ depend on
the order of cutoff $\nmax$ in solving the generator $iG(\omega)$
by the $\omega$-expansion method,
but they themselves have to be calculated non-perturbatively, especially
for $\cG$ by Eq.~(\ref{eq:QPtraTdiagD}) with~(\ref{eq:gdiagD}).
As noticed in the end of \S\ref{sec:SCCsol},
we can apply the SCC method starting from the finite frequency $\omega_0$.
In such a case $\cU_0$ is the transformation at $\omega=\omega_0$,
and $\cG$ and $\cF$ are obtained by expansions
in terms of $(\omega-\omega_0)$; thus,
\beq
  \cU(\omega)^{\rm SCC} =
  \cU_0(\omega_0) \,\cG(\omega-\omega_0) \cF(\omega-\omega_0)
  \quad{\rm if\;started\;at\;}\omega=\omega_0.
\label{eq:QPtraSCCUa}
\eeq
It should be stressed that the transformation~(\ref{eq:QPtraSCCU})
only approximately diagonalize the Hamiltonian
in Eq.~(\ref{eq:diagCSM}) within the $\nmax$ order in the sense
of $\omega$-expansion.  Namely, some parts of the Hamiltonian
corresponding to the terms higher order than $\nmax$ are neglected,
and this is exactly the reason why we can obtain the diabatic
basis, whose negative and positive solutions are non-interacting.

   In the case where the effect of residual interactions is neglected,
i.e. corresponding to the higher order cranking,
we can easily solve the basic equations of the SCC method.
It is useful to present the solution for practical purposes; for example
for the construction of the diabatic quasiparticle basis
for the cranked shell model calculations.
The solutions for $g^{(n)}$ up to the third order are given as follows:
\ben
  g^{(1)}(ij) &=& \frac{1}{E_i + E_{\bar j}} J_x^A(ij),
\label{eq:SCCsolCSM1} \\
  g^{(2)}(ij) &=& \frac{1}{E_i + E_{\bar j}}
   (J_x^B g^{(1)}+g^{(1)} {\bar J}_x^B)_{ij},
\label{eq:SCCsolCSM2} \\
  g^{(3)}(ij) &=& \frac{1}{E_i + E_{\bar j}}
   \bigl[(J_x^B g^{(2)}+g^{(2)} {\bar J}_x^B) \nonumber \\
 &&\quad
   +\frac{1}{3}(J_x^A g^{(1)\T} g^{(1)} + 2 g^{(1)} J_x^{A\T} g^{(1)}
     + g^{(1)} g^{(1)\T} J_x^A) \bigr]_{ij},
\label{eq:SCCsolCSM3}
\een
and the solutions for the rotating quasiparticle
Hamiltonian~(\ref{eq:DiabQPh})$-$(\ref{eq:DiabQPemat}):
\ben
  \epsilon^{\prime(0)}_{ij} &=& \delta_{ij} E_i, \quad
  {\bar \epsilon}^{\prime(0)}_{ij} = \delta_{ij} E_{\bar i},
\label{eq:SCCsolCSMQPemat0} \\
  \epsilon^{\prime(1)}_{ij} &=& -J_x^B(ij), \quad
  {\bar \epsilon}^{\prime(1)}_{ij} = -{\bar J}_x^B(ij),
\label{eq:SCCsolCSMQPemat1} \\
  \epsilon^{\prime(2)}_{ij} &=& \frac{1}{2}
  (J_x^A g^{(1)\T}+g^{(1)}J_x^{A\T})_{ij}, \quad
  {\bar \epsilon}^{\prime(2)}_{ij} = \frac{1}{2}
  (J_x^{A\T} g^{(1)}+g^{(1)\T}J_x^A)_{ij},
\label{eq:SCCsolCSMQPemat2} \\
  \epsilon^{\prime(3)}_{ij} &=& \frac{1}{2}
  (J_x^A g^{(2)\T}+g^{(2)}J_x^{A\T})_{ij}, \quad
  {\bar \epsilon}^{\prime(3)}_{ij} = \frac{1}{2}
  (J_x^{A\T} g^{(2)}+g^{(2)\T}J_x^A)_{ij},
\label{eq:SCCsolCSMQPemat3}
\een
where the quasiparticle energies at the starting frequency are given
in Eq.~(\ref{eq:Ophform}),
and the matrix elements of $J_x$ at the starting frequency are given
as in Eq.~(\ref{eq:OpQform}) with $Q_\rho$ replaced by $J_x$.
If the starting frequency is $\omega=0$, then $E_{\bar i}=E_i$,
and the matrix elements of $J_x$ satisfy the relations, $J_x^{A\T} =- J_x^A$,
${\bar J}_x^B = -J_x^B$, and $J_x^{B\T}=J_x^B$.
The transformation $\cG(\omega)$ is calculated
from Eqs.~(\ref{eq:SCCsolCSM1})$-$(\ref{eq:SCCsolCSM3}), and $\cF(\omega)$
from Eqs.~(\ref{eq:SCCsolCSMQPemat0})$-$(\ref{eq:SCCsolCSMQPemat3}).
It should be mentioned that the selfconsistent mean-field calculation
is in principle possible in combination with the diabatic basis
prescription presented above.

\subsection{ Estimate of the $g$-$s$ interaction}
\label{sec:gsInt}

   Once the diabatic $g$- and $s$-bands states~(\ref{eq:gsWaveFun})
are obtained as functions of $\omega$, one can immediately construct them
as functions of angular momentum $I$, because the $I_x-\omega$ relation has
no singularity, as shown in Fig.~\ref{fig:SCCgsall},
and can easily be inverted:
\beq
   \ket{\phi_g(I)}=\ket{\phi_g(\omega_g(I))}, \quad
   \ket{\phi_s(I)}=\ket{\phi_s(\omega_s(I))},
\label{eq:gsWaveFunI}
\eeq
where $\omega_g(I)$ and $\omega_s(I)$ are
the inverted relations of~(\ref{eq:Iofgs}) with $I_x=I+1/2$.
Physically, one has to consider the coupling problem between them
at a fixed spin value $I$.  It is, however, a difficult problem
because one has to calculate, for example, a matrix element like
$\braketc{\phi_s(I)}{H}{\phi_g(I)}$, which is an overlap
between two different HB states;
they are not orthogonal to each other due to
the difference of the frequencies $\omega_g(I)$ and $\omega_s(I)$.
Although such a calculation is possible by using the Onishi formula
for the overlap of general HB states,\cite{RStext} \ 
it would damage the simple picture of quasiparticle motions
in the rotating frame, and is out of scope of the present investigation.

   Here we assume that the wave functions varies smoothly along
the diabatic rotational bands as functions of spin $I$ or frequency $\omega$,
so that the interband interaction between the $g$- and $s$-bands
can be evaluated at the common frequency by
\beq
  v_{g\mbox{-}s}(I) =
  \braketc{\phi_s(\omega_{gs}(I))}{H}{\phi_g(\omega_{gs}(I))},
\label{eq:IntVgs}
\eeq
where $\omega_{gs}$ is defined by an average of $\omega_s$ and $\omega_g$,
\beq
  \omega_{gs}(I) \equiv \frac{\omega_g(I)+\omega_s(I)}{2}.
\label{eq:Omegags}
\eeq
We note that this quantity corresponds, in a good approximation,
to the crossing frequency $\omega_{\rm c}^{g\mbox{-}s}$
at the crossing angular momentum $I_{\rm c}^{g\mbox{-}s}$,
\beq
   \omega_{gs}(I_{\rm c}^{g\mbox{-}s}) \approx \omega_{\rm c}^{g\mbox{-}s},
\label{eq:Omegac}
\eeq
where $\omega_{\rm c}^{g\mbox{-}s}$ is defined
as a frequency at which the lowest diabatic two quasiparticle
energy vanishes, $E'_1(\omega)+E'_{\bar 1}(\omega) = 0$.
Using the fact that $\ket{\phi_s(\omega)}$ is the two quasiparticle
excited state on $\ket{\phi_g(\omega)}$ (see Eq.~(\ref{eq:gsWaveFun})),
the interaction can be rewritten as
\beq
  v_{g\mbox{-}s}(I) = \omega_{gs}(I)
  \braketc{\phi_s(\omega_{gs}(I))}{J_x}{\phi_g(\omega_{gs}(I))},
\label{eq:IntVgsa}
\eeq
because of the variational principle~(\ref{eq:eqpathF}).
Applying the idea of $\omega$-expansion and taking up to the lowest
order, we have, at the crossing angular momentum $I_{\rm c}^{g\mbox{-}s}$,
\beq
  v_{g\mbox{-}s}(I_{\rm c}^{g\mbox{-}s}) \approx
  \omega_{\rm c}^{g\mbox{-}s}
  \sum_{ij>0}f_{i1}(\omega_{\rm c}^{g\mbox{-}s})
  {\bar f}_{j1}(\omega_{\rm c}^{g\mbox{-}s}) J_x^A(ij),
\label{eq:IntVgsb}
\eeq
where $f_{i1}(\omega)$ and ${\bar f}_{j1}(\omega)$ are the amplitudes
of the diabatic quasiparticle diagonalization~(\ref{eq:eigQPE})
for the lowest $r=\pm i$ quasineutrons,
and should be calculated non-perturbatively with respect to $\omega$.

\begin{figure}
\epsfysize=6.0cm
\centerline{\epsfbox{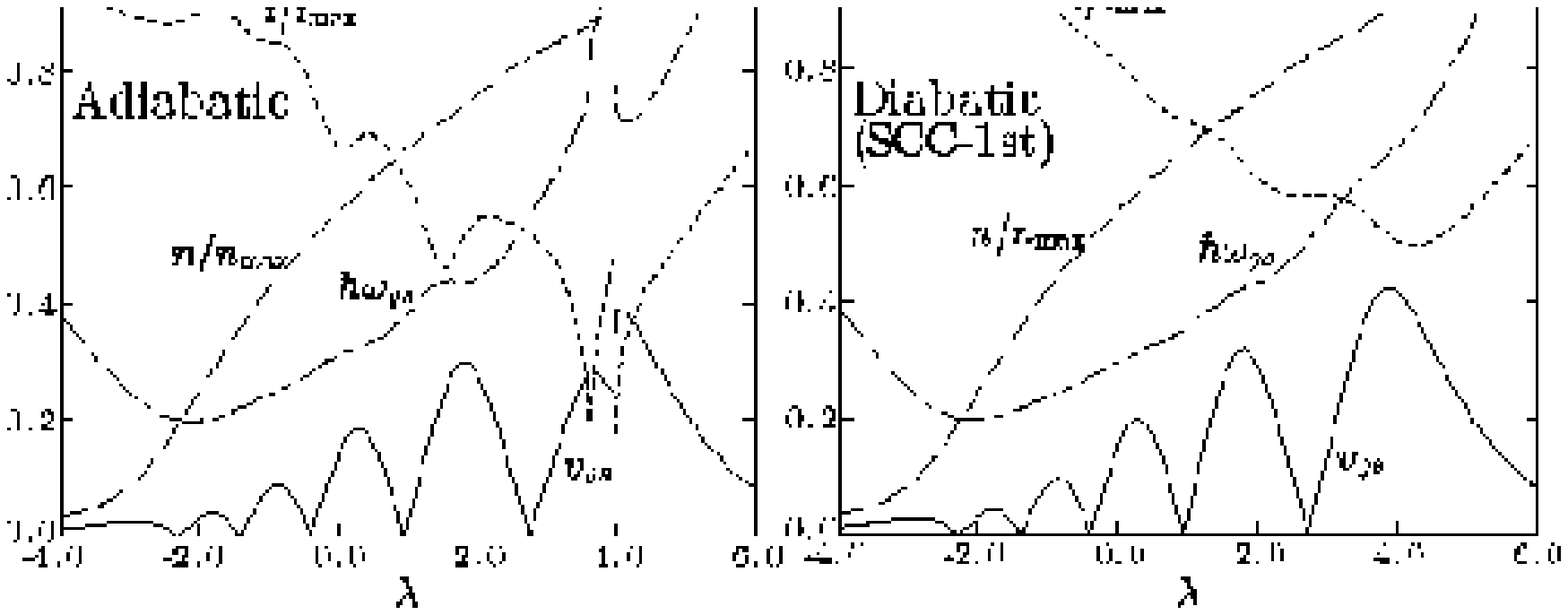}}
\caption{
 The $g$-$s$ interband interaction (solid),
 the crossing frequency $\hbar\omega_{gs}$ (dash-dotted) in MeV,
 the alignment $i$ (dotted), and the expectation value of number 
 operator $n$ (dashed),
 plotted as functions of the chemical potential $\lambda$ in MeV,
 for the $i_{13/2}$ single-$j$ shell model without residual interactions.
 The result of the usual adiabatic cranked shell model
 is displayed in the left panel, while
 that of the diabatic SCC 1st order calculation in the right panel.
 Here the alignment $i$ and the number $n$ is scaled by their maximum
 values, $i_{\rm max}=12\hbar$ and $n_{\rm max}=14$.  The energy unit
 is chosen such that the splitting of the $i_{13/2}$-shell roughly
 reproduces that of a typical well deformed rare-earth nucleus;
 i.e. $\kappa=2.5$ MeV in Eq.~(\protect\ref{eq:singjen}),
 and the constant $\Delta=1.0$ MeV is used.
  }
\label{fig:intsinglj}
\end{figure}

   In Fig.~\ref{fig:intsinglj} (right panel),
we show the result evaluated by using Eq.~(\ref{eq:IntVgsb})
for a simple single-$j$ shell model ($i_{13/2}$)
with a constant monopole-pairing gap and no residual interactions,
in which the single-particle energies are given by
\beq
   e_i = \kappa \,\frac{3m_i^2-j(j+1)}{j(j+1)}\quad (m_i=1/2,\cdots,j),
\label{eq:singjen}
\eeq
with a parameter $\kappa$ describing the nuclear deformation.
In this figure other quantities,
the alignment of the lowest two quasiparticle state,
the number expectation value, and the crossing frequency are also shown
as functions of the chemical potential.
These quantities can also be evaluated in terms of the usual adiabatic cranking
model, and they are also displayed in the left panel.
Note that in the adiabatic cranking model the crossing frequency is
defined as a frequency at which the adiabatic two quasiparticle
energy $E^{\prime({\rm ad})}_1(\omega)+E^{\prime({\rm ad})}_{\bar 1}(\omega)$
becomes the minimum, and the interband interaction is identified
as the half of its minimum value.\cite{BF79} \ 
As is well known,\cite{BHM78} \ the $g$-$s$ interaction oscillates
as a function of the chemical potential,
and both the absolute values and the oscillating behavior of the
result of calculation roughly agree with the experimental findings.
Comparing two calculations, the interband interaction~(\ref{eq:IntVgsb})
seems to give a possible microscopic estimate based on the diabatic
description of the $g$- and $s$-bands.
We would like to stress, however, that its derivation is not very sound.
It is an important future problem 
to derive the coupling matrix element on a more sound ground.

% **********
% * Figure
% **********

% ***********************************
% * SECTION 5 (CONCLUSION)
% ***********************************

\section{ Concluding remarks }

    In this paper, we have formulated the SCC method for the nuclear
collective rotation.  By using the rotational frequency expansion
rather than the angular momentum expansion, we have applied it
to the description of the $g$- and $s$-bands successfully.
The systematic calculation gives surprisingly good agreements
with experimental data for both rotational bands.
It has been demonstrated that the resultant quasiparticle states
develop diabatically as functions of the rotational frequency;
i.e. the negative and positive energy levels do not interact with each other.
Although the formulation is mathematically equivalent to the selfconsistent
cranking model, the cutoff of the $\omega$-expansion results in
the diabatic levels and its mechanism is also discussed.
The perturbative $\omega$-expansion is, however, inadequate to use
the resultant quasiparticle basis states as a complete set.
We have then presented a method to construct the diabatic quasiparticle
basis set, which rigorously satisfies the orthonormality condition 
and can be safely used for the next step calculation, e.g.
the RPA formalism for collective vibrations at high-spin.

    In order to obtain a good overall description of the rotational band
for nuclei in the rare-earth region, we have investigated the best
possible form of residual quadrupole-pairing interactions.  
It is found that the double-stretched 
form factor is essential for reproducing the even-odd mass difference
and the moment of inertia simultaneously.

    Since the calculated $g$- and $s$-bands in our formulation are
diabatic rotational bands, the interband interaction between them
should be taken into account for their complete descriptions.
As in any other mean-field model, however, the wave function obtained
in our formalism is a wave packet 
with respect to the angular momentum variable.
Therefore, it is not apparent how to evaluate the interband
interaction from microscopic point of view.
We have presented a possible estimate of the interaction,
which leads to a value similar to that estimated by the level repulsion
in the adiabatic cranking model.
Further investigations are still necessary to give a definite conclusion
to this problem.

\section*{ Acknowledgments }

   It is a great pleasure for us to publish this paper on the occasion 
of the 70th birthday of Professor Marumori, to whom we are deeply indebted 
for guiding us to many-body theory of nuclear collective motions.
This work is supported in part by the Grant-in-Aid for
Scientific Research from the Japan Ministry of Education,
Science and Culture (No. 12640281).

% ***********************************
% * REFERENCES
% ***********************************


\begin{thebibliography}{99}

%%% ********** high spin review **********
\bibitem{Hama85} I.~Hamamoto,
{\it Treatise on Heavy-Ion Science}, Vol. 3, ed. D.~A.~Bromly
(Prenum Press, 1985), p. 313.
\bibitem{Szy83} Z.~Szymansky,
{\it Fast Nuclear Rotation}, Clarendon Press, 1983.

%%% ********** band-crossing difficulty **********
\bibitem{Hama76} I.~Hamamoto,
  Nucl. Phys. {\bf A271} (1976), 15.
\bibitem{MG78} E.~R.~Marshalek and A.~L.~Goodman,
  Nucl. Phys. {\bf A294} (1978), 92.
\bibitem{BF79a} R.~Bengtsson and S.~Frauendorf,
  Nucl. Phys. {\bf A314} (1979), 27.
\bibitem{Fra82} S.~Frauendorf, in {\it Nuclear Physics},
ed. C.~H.~Dasso, R.~A.~Broglia and A.~Winther (North-Holland, 1982),
p.111.

%%% ********** our paper **********
\bibitem{SM83} Y.~R.~Shimizu and K.~Matsuyanagi,
  Prog. Theor. Phys. {\bf 70} (1983), 144; {\bf 72} (1984), 799.

%%% ********** other method of diabatic band **********
\bibitem{Ben89} T.~Bengtsson,
  Nucl. Phys. {\bf A496} (1989), 56.
\bibitem{MDV97} M.~Matsuo, T.~D\o ssing, E.~Vigezzi,
  R.~A.~Broglia and K.~Yoshida,
  Nucl. Phys. {\bf A617} (1997), 1.

%%% ********** SCCM original **********
\bibitem{MMSK80} T.~Marumori, T.~Maskawa, F.~Sakata and A.~Kuriyama,
  Prog. Theor. Phys. {\bf 64} (1980), 1294.

%%% ********** P+QQ model review **********
\bibitem{BS69} D.~R.~Bes and R.~A.~Sorensen,
{\it Advances in Nuclear Physics}, Vol. 2, (Prenum Press, 1969), p.129.

%%% ********** pairing Gap systematics **********
\bibitem{NP61} S.~G.~Nilsson and O.~Prior,
  Mat. Fys. Medd. Dan. Vid. Selsk. {\bf 32} (1961), no.~16.

%%% ********** Skyrme force **********
\bibitem{VB72} D.~Vautherin and D.~M.~Brink,
Phys. Rev. {\bf C3} (1972), 626.

%%% ********** Strutinsky-Nilsson method **********
\bibitem{Brack72} M.~Brack, J.~Damgaard, A.~S.~Jensen, H.~C.~Pauli,
V.~M.~Strutinsky and C.~Y.~Wong, Rev. Mod. Phys. {\bf 44} (1972), 320.
\bibitem{RNS78} I.~Ragnarsson, S.~G.~Nilsson and R.~K.~Sheline,
Phys. Rep. {\bf 45} (1978), 1.
\bibitem{NR95} S.~G.~Nilsson and I.~Ragnarsson, 
{\it Shapes and Shells in Nuclear Structure}, 
Cambridge University Press, 1995.

%%% ********** Double-stretched QQ **********
\bibitem{SK89} H.~Sakamoto and T.~Kishimoto,
  Nucl. Phys. {\bf A501} (1989), 205; 242.
\bibitem{Kis75} T.~Kishimoto et al.,
  Phys. Rev. Lett. {\bf 35} (1975), 552;
  T.~Kishimoto, {\it Proc. 1980 RCNP Int. Symp. Highly Exited States
  in Nuclear Reactions, Osaka}, p.145.
\bibitem{Mar84} E.~R.~Marshalek,
  Phys. Rev.  {\bf 29} (1984), 640.

%%% ********** Iso-Stretching of QQ **********
\bibitem{BK68} M.~Baranger, K.~Kumar,
  Nucl. Phys. {\bf A110} (1968), 490.
\bibitem{Sak93} H.~Sakamoto,
  Nucl. Phys. {\bf A557} (1993), 583c.

%%% ********** Q-pairing **********
\bibitem{Hama74} I.~Hamamoto,
  Nucl. Phys. {\bf A232} (1974), 445.
\bibitem{Dieb84} M.~Diebel,
  Nucl. Phys. {\bf A419} (1984), 353.
\bibitem{Garr82} J.~D.~Garrett et al.,
  Phys. Lett. {\bf B118} (1982), 297.

%%% ********** Double-stretched Q-pairing **********
\bibitem{SK90} H.~Sakamoto and T.~Kishimoto,
  Phys. Lett. {\bf B245} (1990), 321.
\bibitem{KSKK96} T.~Kubo, H.~Sakamoto, T.~Kammuri and T.~Kishimoto,
  Phys. Rev. {\bf C54} (1996), 2331.
\bibitem{SW94} W.~Satu{\l}a and R.~Wyss,
  Phys. Rev. {\bf C50} (1994), 2888.

%%% ********** texts **********
\bibitem{BMtextI} A.~Bohr and B.~R.~Mottelson,
{\it Nuclear Structure}, Vol.~I, (W.~A.~Benjamin Inc., 1969).
\bibitem{BMtextII} A.~Bohr and B.~R.~Mottelson,
{\it Nuclear Structure}, Vol.~II, (W.~A.~Benjamin Inc., 1975).

%%% ********** Harris formula **********
\bibitem{Har65} S.~M.~Harris,
Phys. Rev. {\bf 138} (1965), B509.

%%% ********** Nilsson parameters *********
\bibitem{BR85}
T.~Bengtsson and I.~Ragnarsson,
Nucl. Phys. {\bf A436} (1985) 14.

%%% ********** exp.data etc. ********
\bibitem{AW93} G.~Audi and A.~H.~Wapstra,
  Nucl. Phys. {\bf A565} (1993), 1.

\bibitem{Sakai84} M.~Sakai,
Atomic and Nuclear Data Tables, {\bf 31} (1984), 399.

\bibitem{ENSDF} Evaluated Nuclear Structure Data File,
http://isotopes.lbl.gov/isotopes/vuensdf.html.

%%% ********** texts **********
\bibitem{RStext} P.~Ring and P.~Schuck,
{\it The Nuclear Many-Body Problem},
(Springer-Verlag, 1980).

%%% ********** our paper **********
\bibitem{SM85} Y.~R.~Shimizu and K.~Matsuyanagi,
  Prog. Theor. Phys. {\bf 74} (1985), 1346.

%%% ********** SCCM for rotation **********
\bibitem{TMS91} J.~Terasaki, T.~Marumori and F.~Sakata,
  Prog. Theor. Phys. {\bf 85} (1991), 1235.
\bibitem{Tera92} J.~Terasaki,
  Prog. Theor. Phys. {\bf 88} (1992), 529.
\bibitem{Tera94} J.~Terasaki,
  Prog. Theor. Phys. {\bf 92} (1994), 535.

%%% ********** cranked shell model *********
\bibitem{BF79} R.~Bengtsson and S.~Frauendorf,
  Nucl. Phys. {\bf A327} (1979), 139.

%%% ********** SCC number treatment **********
\bibitem{Mat86} M.~Matsuo,
  Prog. Theor. Phys. {\bf 76} (1986), 372.

%%% ********** SCC gamma vibration **********
\bibitem{Mat92} M.~Matsuo, 
  Springer Proceedings in Physics, Vol. 58,
  {\it New Trends in Nuclear Collective Dynamics},
  ed. Y.~Abe, H.~Horiuchi, K.~Matsuyanagi (Springer-Verlag, 1992),
  p.219.
\bibitem{Mat84} M.~Matsuo,
  Prog. Theor. Phys. {\bf 72} (1984), 666.
\bibitem{MMgamma} M.~Matsuo and K.~Matsuyanagi,
  Prog. Theor. Phys. {\bf 74} (1985), 1227; {\bf 76} (1986), 93;
  {\bf 78} (1987), 591.
\bibitem{MSMgamma} M.~Matsuo, Y.~R.~Shimizu and K.~Matsuyanagi,
  Proc. of the Niels Bohr Centennial Conf. on Nuclear Structure,
  ed. R.~Broglia, G.~Hagemann, B.~Herskind (North-Holland, 1985),
  p.161.

%%% ********** SCC  transitional nuclei **********
\bibitem{TYT89} K.~Takada, K.~Yamada and H.~Tsukuma,
  Nucl. Phys. {\bf A496} (1989) 224.
\bibitem{YTT89} K.~Yamada, K.~Takada and H.~Tsukuma,
  Nucl. Phys. {\bf A496} (1989) 239.
\bibitem{YT89} K.~Yamada and K.~Takada,
  Nucl. Phys. {\bf A503} (1989) 53.
\bibitem{Aiba90} H.~Aiba,
  Prog. Theor. Phys. {\bf 83} (1990), 358; {\bf 84} (1990), 908.
\bibitem{Yamada91} K.~Yamada,
  Prog. Theor. Phys. {\bf 85} (1991), 805.
\bibitem{Yamada93} K.~Yamada,
  Prog. Theor. Phys. {\bf 89} (1993), 995.

%%% ********** our paper **********
\bibitem{MSM88} M.~Matsuzaki, Y.~R.~Shimizu and K.~Matsuyanagi,
  Prog. Theor. Phys. {\bf 79} (1988), 836.

%%% ********** Tanaka Suekane **********
\bibitem{TS81} Y.~Tanaka and S.~Suekane,
  Prog. Theor. Phys. {\bf 66} (1981), 1639.

%%% ********** exp.data etc. ********
\bibitem{Loeb70} K.~E.~G.~L\"obner, M.~Vetter and V.~H\"onig,
Nculear Data Tables {\bf A7} (1970), 495.

%%% ********** Phenomenological Gap  **********
\bibitem{WSNJ90} R.~Wyss, W.~Satu{\l}a, W.~Nazarewicz, and A.~Johnson,
  Nucl. Phys. {\bf A511} (1990), 324.

%%% ********** band termination  **********
\bibitem{AFLR99} A.~V.~Afanasjev, D.~B.~Fossan, G.~J.~Lane and I.~Ragnarsson,
  Phys. Rep. {\bf 322} (1999), 1.

%%% ********** RPA at high spin *********
\bibitem{Mar77} E.~R.~Marshalek,
  Nucl. Phys. {\bf A275} (1977), 416; {\bf A331} (1979), 429.
\bibitem{JM79} D.~Janssen and I.~N.~Mikhailov,
  Nucl. Phys. {\bf A318} (1979), 390.
\bibitem{EMR80} J.~I.~Egido, H.~J.~Mang and P.~Ring, 
  Nucl. Phys. {\bf A339} (1980), 390.
\bibitem{Zel80} V.~G.~Zelevinsky, 
  Nucl. Phys. {\bf A344} (1980), 109.

%%% ********** gs interaction **********
\bibitem{BHM78} R.~Bengtsson, I.~Hamamoto and B.~Mottelson,
Phys. Lett. {\bf 73B} (1978), 259.

\end{thebibliography}
\end{document}